\documentclass[journal, 10pt]{IEEEtran}
\IEEEoverridecommandlockouts
\usepackage[subrefformat=parens]{subcaption}
\usepackage{amsmath,graphicx}
\usepackage{amsmath,amssymb,amsfonts}
\usepackage{cite,soul}
\usepackage{url}
\usepackage{bm}
\usepackage{comment}
\usepackage{tabularx}
\DeclareMathOperator*{\argmax}{argmax}
\usepackage[usenames,dvipsnames]{color}
\usepackage[linesnumbered,vlined,ruled]{algorithm2e}
\usepackage{algpseudocode}

\newenvironment{algocolor}{%
   \setlength{\parindent}{0pt}
   \itshape
   \color{blue}
}{}

\title{Process-and-Forward: Deep Joint Source-Channel Coding Over Cooperative Relay Networks}
\author{Chenghong Bian,~\IEEEmembership{Graduate Student Member,~IEEE}, Yulin Shao,~\IEEEmembership{Member,~IEEE}, Haotian Wu,~\IEEEmembership{Graduate Student Member,~IEEE}, Emre Ozfatura,~\IEEEmembership{Member,~IEEE}, Deniz G{\"u}nd{\"u}z,~\IEEEmembership{Fellow,~IEEE}
\thanks{C. Bian, H. Wu, E. Ozfatura, and D. G{\"u}nd{\"u}z are with the Department of Electrical and Electronic Engineering, Imperial College London, London SW7 2AZ, U.K. (e-mails: \{c.bian22, haotian.wu17, m.ozfatura, d.gunduz\}@imperial.ac.uk).}
\thanks{Y. Shao is with the State Key Laboratory of Internet of Things for Smart City and the Department of Electrical and Computer Engineering, University of Macau, Macau S.A.R (e-mail: ylshao@um.edu.mo). \textit{(Corresponding author: Yulin Shao)}}
\thanks{This work received funding in part from the UKRI for the projects AI-R (ERC Consolidator Grant, EP/X030806/1) and INFORMED-AI (EP/Y028732/1), the SNS JU project 6G-GOALS under the EU's Horizon program (Grant Agreement No. 101139232), and the Science and Technology Development Fund, Macao (Project no.: 0068/2023/RIB3 and 0062/2024/RIA1).}
\thanks{For the purpose of open access, the authors have applied a Creative Commons Attribution (CC BY) license to any author accepted manuscript version arising.}
\thanks{This paper was presented in part at the IEEE International Conference on Machine Learning for Communication and Networking, 2024 \cite{relay_jscc}.}
}

\begin{document}

\maketitle

\thispagestyle{empty}
\begin{abstract}
We introduce deep joint source-channel coding (DeepJSCC) schemes for image transmission over cooperative relay channels. The relay either amplifies-and-forwards its received signal, called DeepJSCC-AF, or leverages neural networks to extract relevant features from its received signal, called DeepJSCC-PF (Process-and-Forward). We consider both half- and full-duplex relays, and propose a novel transformer-based model at the relay. For a half-duplex relay, it is shown that the proposed scheme learns to generate correlated signals at the relay and source to obtain beamforming gains. In the full-duplex case, we introduce a novel {block-based} transmission strategy, in which the source transmits in blocks, and the relay updates its knowledge about the input signal after each block and generates its own signal. To enhance practicality, a single transformer-based model is used at the relay at each block, together with an adaptive transmission module, which allows the model to seamlessly adapt to different channel qualities {and the transmission powers}. Simulation results demonstrate the superior performance of DeepJSCC-PF compared to the state-of-the-art BPG image compression algorithm operating at the maximum achievable rate of conventional decode-and-forward and compress-and-forward protocols, in both half- and full-duplex relay scenarios {over AWGN and Rayleigh fading channels}.
\end{abstract}

\begin{IEEEkeywords} Deep joint source-channel coding, cooperative relay networks, decode-and-forward, {deep learning-based transceiver design.} \end{IEEEkeywords}

\section{Introduction}\label{sec:intro}
\begin{figure}[t]
     \centering
     \begin{subfigure}{\columnwidth}
         \centering
         \includegraphics[width=0.75\columnwidth]{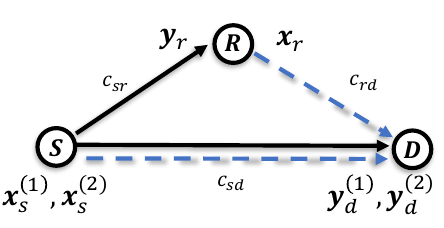}
         \caption{Half-duplex relaying: The solid arrows denote the transmission in the \textit{relay-receive} period while the dashed ones represent that in the \textit{relay-transmit} period.}
         \label{fig:system_model:a}
     \end{subfigure}     
     \begin{subfigure}{\columnwidth}
         \centering
         \includegraphics[width=0.7\columnwidth]{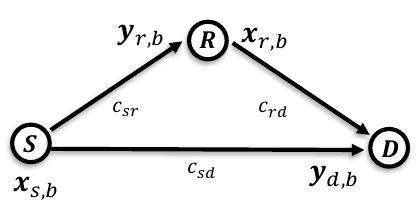}
         \caption{Full-duplex relaying: The relay can receive and transmit simultaneously.}
         \label{fig:system_model:b}
     \end{subfigure}

  \caption{Illustrations of the half-duplex and full-duplex relay channels.}
  \label{fig:system_model}
\end{figure}

Cooperative communications empower nodes within a network to harness their neighbors' resources, enhancing spectral efficiency and fortifying the network against channel fading \cite{usr_coop}.
{The most fundamental cooperative communication model is the \textit{relay channel}, comprising three nodes: the source, the relay, and the destination, as depicted in Fig. \ref{fig:system_model}.}
The source transmits its message to both the relay and destination. The relay processes its received signal and transmits it to the destination, while the destination tries to recover the original message by combining the signals received from the source and relay. Three classical relaying schemes are commonly employed: amplify-and-forward (AF), decode-and-forward (DF), and compress-and-forward (CF) \cite{relay_capacity0, relay_capacity, relay_capacity2, nit, CF, tit_cf, orthogonal_relay}.
In AF, the relay straightforwardly scales and forwards its received signal to the destination, which suffers from noise forwarding. In DF, the relay decodes the received message before re-encoding and forwarding it. While DF mitigates the noise forwarding issue, it faces limitations when the source-to-relay channel quality is poor. 
CF, on the other hand, adopts a different approach, having the relay compress its received signal using Wyner-Ziv source coding \cite{wyner-ziv}, while treating the destination's received signal as side information. Despite many efforts, the capacity of a general relay channel {remains an open problem to this date.}

Remarkably, it has been demonstrated in \cite{titjscc_relay} that separate compression followed by cooperative channel coding is optimal under infinite source and channel block lengths, even though the capacity of the relay channel cannot be computed accurately. However, in practical finite block length regimes, joint source-channel coding (JSCC) often outperforms the separation approach, though there is limited research on JSCC applied to relay channels. The study in \cite{titjscc_relay} delves into JSCC over cooperative relay networks from an information-theoretic perspective, {and proposes various achievable schemes}. Additionally, JSCC for cooperative multimedia source transmission has been explored in \cite{coop_multimedia}, {where separate codes for compression and error correction are optimized jointly to bolster resilience against channel variations.}

More recently, deep learning (DL) has made significant strides in addressing various communication challenges, particularly in the realm of JSCC \cite{deepjscc,deepjscc_semantic,VIT,vit_vs_cnn, deepjscc_mimo_twc, bupt_vit,summary_vit,jsccofdm,st_jscc,dtat2022}. The DeepJSCC scheme introduced in \cite{deepjscc} has demonstrated superiority over conventional digital approaches, combining state-of-the-art compression techniques with nearly optimal channel codes for image transmission across AWGN and Rayleigh fading channels. Moreover, it exhibits graceful degradation as channel quality weakens. DeepJSCC framework \cite{deepjscc_semantic} has already extended to many different wireless communication channels  such as OFDM \cite{jsccofdm}, MIMO \cite{st_jscc, deepjscc_mimo_twc} channels for image transmission. {While the initial studies \cite{deepjscc, deepjscc_semantic} relied mainly on convolutional architectures, recent studies have shown that vision transformers (ViTs) \cite{VIT} can achieve superior results \cite{vit_vs_cnn, deepjscc_mimo_twc, bupt_vit, summary_vit}}.
{The DeepJSCC framework has also been applied to other signals, such as video \cite{deepwive, bupt_video}, audio \cite{semantic_speech}, and point cloud sources \cite{sept}, highlighting it as an effective tool for a large variety of applications.}

Despite these notable advances in applying DeepJSCC in critical scenarios, prior work has primarily concentrated on point-to-point channels, leaving the more intricate multi-user scenarios largely unexplored. Notably, the authors in \cite{mac_jscc} consider DeepJSCC over a multiple access channel. 
DeepJSCC is also applied to image transmission over the broadcast \cite{wu2023fusionbased} and the multi-hop relay channels \cite{multihop_jscc, IOTJ_2024, hybrid_jscc}. However, when it comes to cooperative communications, there appears to be a significant gap in research. The most relevant work to the current paper is \cite{semantic_relay_speech}, which considers the three-node cooperative relay channel; {yet it is specifically designed for speech transmission, and focus on the energy efficiency.}
Our previous work \cite{relay_jscc} is the first to investigate image transmission over the half-duplex relay channel. However, the scenario studied in \cite{relay_jscc} is limited as it assumes that the source node keeps silent during the \textit{relay-transmit} period. 

In this paper, we introduce the first DeepJSCC framework tailored for both the half-duplex and full-duplex relay channels. Within this framework, we present two distinctive relaying protocols: DeepJSCC-AF and DeepJSCC-PF. These protocols draw parallels with the classical relaying schemes and are designed to operate effectively in both half-duplex and full-duplex relay scenarios.
In the DeepJSCC-AF protocol, the relay simply amplifies its received signal while adhering to power constraints, while the encoder and decoder networks are trained jointly to benefit from the signal forwarded by the relay. On the other hand, DeepJSCC-PF involves signal processing at the relay using a deep neural network (DNN). {The relay DNN is} jointly optimized with the DNNs at the source and destination nodes to achieve superior {end-to-end} reconstruction performance. The inspiration for the DeepJSCC-PF protocol is rooted in the principles of the traditional DF and CF protocols. Note that, {in PF}, there is no decoding {at the relay} in the strict sense since DeepJSCC does not rely on digital encoding of information; and hence, the source signal cannot be decoded accurately, and the relay can only estimate the source signal.  DeepJSCC-PF scheme also resembles the classical CF scheme as the received signal is compressed and forwarded, albeit using a {JSCC} scheme, instead of separate compression followed by channel coding. Moreover, DeepJSCC-PF also learns to exploit the structure of the received signal, while in CF, the received signal is simply treated as an  independent and identically distributed source sequence.  
We meticulously search for the optimal parameters of the proposed scheme (time division and power allocation) to enhance its performance in the half-duplex relay, while we apply a block-based transmission strategy inspired by the well-known block Markov coding in the full-duplex scenario.

Our dedicated experiments unequivocally demonstrate the effectiveness of the proposed DeepJSCC-PF protocol in both half-duplex and full-duplex scenarios. To enhance the practicality of these DeepJSCC schemes, particularly in the context of full-duplex relay, we move beyond training separate models for different blocks and varying channel/link conditions. Instead, we propose a single adaptive transformer-based encoder that refines its knowledge of the source after each block, and leverages link conditions as side information to attain reconstruction performance on par with individually trained models over different network states.

Next, we summarize the main contributions of this paper:

\begin{itemize}
    \item We present a novel DeepJSCC framework for both the half-duplex and full-duplex relay channels, and propose the novel DeepJSCC-AF and DeepJSCC-PF coding schemes. Central to our framework is a transformer-based coding architecture, inspired by the ViT, which parameterizes the encoding and decoding functions across the source, relay, and destination nodes. This approach ensures outstanding reconstruction performance in cooperative relay networks, highlighting the great potential of the transformer architecture for practical multi-user code design for future communication networks. 
    \item Our DeepJSCC-PF protocol {in the full-duplex mode} draws inspiration from the block Markov coding (BMC) approach {originally conceived for the DF scheme in \cite{relay_capacity0}}. This scheme pioneers a practical block-based cooperative coding scheme within the JSCC paradigm, offering a fresh perspective on combining block coding with DL for enhanced communication efficiency. We design a matching transformer-based processing module at the relay, which augments its input after each block, and generates the channel input based on relay's current knowledge of the source signal.
    \item {Building upon the SNR-adaptive modules in the literature \cite{xu2021wireless, deepjscclpp}, we introduce a link adaptation (LA) module, specifically designed for cooperative relay networks, which enables dynamic adaptation to varying link qualities and transmit powers in a full-duplex relay scenario. The LA module allows a single adaptive model to match the reconstruction performance of multiple models trained for specific conditions, significantly enhancing operational efficiency.}
    \item Through rigorous numerical experiments in both half-duplex and full-duplex relay configurations {and over both static and Rayleigh fading channels}, we demonstrate the superior performance of the proposed DeepJSCC schemes over existing digital baselines, which employ the BPG image compression algorithm and communicate at the maximum rate achievable by the {DF or CF schemes.}
    Our findings reveal that {the proposed} DeepJSCC schemes not only surpass these baselines in terms of performance but also effectively address the cliff and leveling effects. {Finally, the complexity of the ViT encoding and decoding processes are analyzed for a comprehensive understanding of the proposed scheme.}
\end{itemize}

{\it Notations:} Throughout the paper, normal-face letters (e.g., $x$) represent scalars, while uppercase letters (e.g., $X$) represent random variables. Matrices and vectors are denoted by bold {upper} and {lower} case letters (e.g., $\bm{X}$ and $\bm{x}$), respectively.   {A set is denoted by the double stroke font, e.g., $\mathbb{S}$}. Transpose and Hermitian operators are denoted by $(\cdot)^\top$, $(\cdot)^\dagger$, respectively. We utilize $\bm{x}_{1:b}$ to index the first $b$ elements of a vector $\bm{x}$. {$\text{arg}(h)$ represents the phase of the complex variable, $h$}. {Finally, $\|\bm{S}\|_F$ and $\|\bm{S}\|_{\infty}$ denote the Frobneous and  $\ell$-$\infty$ norm of the matrix $\bm{S}$.}

\section{System model}\label{sec:system_model}

We consider a classical relay channel model consisting of a source node $\mathrm{S}$, a destination node $\mathrm{D}$, and a relay node $\mathrm{R}$, as illustrated in Fig.~\ref{fig:system_model}. The goal is to deliver an image $\bm{S} \in \mathbb{R}^{C\times H \times W}$ from $\mathrm{S}$ to $\mathrm{D}$ with the help of relay $\mathrm{R}$, where $C$, $H$, $W$ denote the number of color channels, the height and width of the image, respectively. The relay node $\mathrm{R}$ can operate in either the half- or the full-duplex mode.

\subsection{Half-Duplex Relaying}
In the half-duplex mode, the relay cannot receive and transmit at the same time, as shown in Fig. \ref{fig:system_model:a}.
Thus, the transmission is divided into two periods \cite{relay_capacity}: the \textit{relay-receive} and \textit{relay-transmit} periods, occupying $\alpha$ and $(1-\alpha)$ proportion of the overall transmission duration, respectively.
The parameter $\alpha$ is termed the ``time-division variable'', signifying its role in determining the temporal allocation of relay operations.

At the beginning of transmission, the source node $\mathrm{S}$ encodes the image $\bm{S}$ into a channel codeword $\bm{x}_s \in \mathbb{C}^k$:
\begin{equation}
\bm{x}_s = f_s(\bm{S}),
\end{equation}
where $f_s(\cdot): \mathbb{R}^{C\times H \times W} \rightarrow \mathbb{C}^k$ is an encoding function, and $\bm{x}_s$  is subject to an average power constraint:
\begin{equation}
\frac{1}{k}\|\bm{x}_s \|^2_2 \leq P_s.
\label{eq:power_const}
\end{equation}

The codeword $\bm{x}_s$ can be partitioned into two parts: $\bm{x}_s = \left[\bm{x}_s^{(1)} ~ \bm{x}_s^{(2)} \right]^T$ with $\bm{x}_s^{(1)} \in \mathbb{C}^{\alpha k}$ for the \textit{relay-receive} period and $\bm{x}_s^{(2)} \in \mathbb{C}^{(1-\alpha) k}$ for the \textit{relay-transmit} period.

The signals received at $\mathrm{R}$ and $\mathrm{D}$ in the relay-receive period are denoted by $\bm{y}_{r}$ and $\bm{y}_{d}^{(1)}$, respectively, and given by
\begin{equation}
\label{equ:relay-receive}
    \bm{y}_{r} = c_{sr}\bm{x}_s^{(1)} + \bm{n}_{r},
\end{equation}
\begin{equation}
\label{equ:destination-receive}
    \bm{y}_{d}^{(1)} = c_{sd}\bm{x}_s^{(1)} + \bm{n}_{d}^{(1)},
\end{equation}
where $c_{sr}$, $c_{sd}$ are real constants governed by the transmission distances of the $\mathrm{S}$-$\mathrm{R}$ and $\mathrm{S}$-$\mathrm{D}$ links, respectively; $\bm{n}_{r}$ and $ \bm{n}_{d}^{(1)}$ denote the independent complex additive white Gaussian noise (AWGN) terms, and without loss of generalizability, we assume $\bm{n}_{r}, \bm{n}_{d}^{(1)} \sim \mathcal{CN}(\bm{0},\bm{I}_{\alpha k})$, where $\bm{I}_{\alpha k}$ denotes an identity matrix with dimension $\alpha k\times \alpha k$. 

Upon receiving $\bm{y}_{r}$, the relay re-encodes it by
\begin{equation}
\bm{x}_r= f_r(\bm{y}_{r}),
\end{equation}
where $f_r(\cdot): \mathbb{C}^{\alpha k} \rightarrow \mathbb{C}^{(1-\alpha) k}$ is the re-encoding function, and $\bm{x}_r \in \mathbb{C}^{(1-\alpha) k}$ is subject to a power constraint:
\begin{equation}
\frac{1}{k}\|\bm{x}_r\|^2_2 \leq P_r.    
\end{equation}
This power constraint ensures that the total energy transmitted by the relay is the same for different $\alpha$ values.


The signal received at the destination in the relay-transmit period is given by:
\begin{equation}
    \bm{y}_{d}^{(2)} = c_{rd}\bm{x}_r + c_{sd}\bm{x}_s^{(2)} + \bm{n}_{d}^{(2)},
    \label{equ:relay-transmit}
\end{equation}
where $c_{rd}$ is the real channel gain for the $\mathrm{R}$-$\mathrm{D}$ link and each element in $\bm{n}_{d}^{(2)}$ is independent and follows a complex Gaussian distribution with zero mean and unit variance.

Given the received signal $\bm{y}_d = \left[ \bm{y}_d^{(1)} ~~ \bm{y}_d^{(2)} \right]^T \in \mathbb{C}^k$ over the two periods, the destination aims to reconstruct the image using a decoding function $g(\cdot): \mathbb{C}^k \rightarrow \mathbb{R}^{C\times H \times W}$. The reconstructed image is given by $\hat{\bm{S}} = g(\bm{y}_{d})$.


\subsection{Full-Duplex Relaying}\label{sec:sec_IIB}
In contrast to half-duplex replaying, the relay can receive and transmit at the same time in the full-duplex mode. Following \cite{LDLC, relay_capacity0}, we consider block transmission: the data transmissions among $\mathrm{S}$, $\mathrm{R}$, and $\mathrm{D}$ occur in a block-wise fashion, and each block transmission occupies the same time.

Source $\mathrm{S}$ encodes the image to $\bm{x}_s = f_s(\bm{S})$ and equally partitions $\bm{x}_s$ to $B$ blocks, yielding $\bm{x}_s = [\bm{x}_{s,1}^\top, \ldots, \bm{x}_{s,B}^\top]^\top$, where $\bm{x}_{s,b} \in \mathbb{C}^{{k}/{B}}$. The same power constraint in \eqref{eq:power_const} is also imposed on $\bm{x}_s$. 

The encoded blocks will be transmitted in $B$ slots. In the $b$-th time slot, the source transmits $\bm{x}_{s,b}$ to both the relay and the destination. At the relay, the received signal $\bm{y}_{r,b}$ can be written as
\begin{equation}
\bm{y}_{r,b} = c_{sr}\bm{x}_{s,b} + \bm{n}_{r,b},
\end{equation}
where $\bm{n}_{r,b} \sim \mathcal{CN}({\bm{0}, \bm{I}_{{k}/{B}}})$.
For the same time slot, the relay node generates and transmits $\bm{x}_{r,b}, b>1$, based on all the previously received signals,
\begin{equation}
\bm{x}_{r,b} = f_r(\bm{y}_{r,1:(b-1)}),
\end{equation}
where $ \bm{y}_{r,1:(b-1)} \triangleq \{\bm{y}_{r,1}, \cdots, \bm{y}_{r,(b-1)}\}$. \footnote{For $b = 1$, since the relay has not received any information from the source, it keeps silent to save energy.}

Denote by $\bm{x}_r = [\bm{x}_{r,1}^\top, \cdots, \bm{x}_{r,B}^\top]^\top$ the signals transmitted by the relay, which satisfy the power constraint $\frac{1}{k}\mathbb{E}(\bm{x}_r^\dagger \bm{x}_r) \leq P_r$.

The $b$-th transmitted block from the source and the relay are superimposed at the destination:
\begin{align}
    \bm{y}_{d,b} = c_{sd}\bm{x}_{s,b} + c_{rd}\bm{x}_{r,b} + \bm{n}_{d,b},
    \label{eq:fd_receiver}
\end{align}
where $\bm{n}_{d,b} \sim \mathcal{CN}({\bm{0}, \bm{I}_{{k}/{B}}})$. After collecting all the received blocks, $\bm{y}_{d} \triangleq [\bm{y}_{d,1}^\top, \cdots, \bm{y}_{d,B}^\top]^\top; \bm{y}_{d} \in \mathbb{C}^{k}$, a decoding function, $g_d(\cdot)$ is used to reconstruct the original image. {We have  $\hat{\bm{S}} = g_d(\bm{y}_d)$ as the reconstruction.} 
The peak signal-to-noise ratio (PSNR) and the structural
similarity index (SSIM) are used to evaluate the reconstruction quality. The PSNR is defined as:
\begin{align}
    {\text{PSNR} = 10 \log_{10} \left(\frac{\|\bm{S}\|^2_{\infty}}{\frac{1}{M} \|\bm{S} - \hat{\bm{S}}\|_F^2} \right),}
    \label{eq:psnr}
\end{align}
where $M \triangleq CHW$ denotes the total number of pixels in the image $\bm{S}$. {The SSIM is defined as:}
\begin{equation}
    \text{SSIM} = \frac{(2\mu_s\mu_{\hat{s}}+c_1)(2\sigma_{s\hat{s}}+c_2)}{(\mu^2_s + \mu^2_{\hat{s}} + c_1)(\sigma^2_s + \sigma^2_{\hat{s}} + c_2)},
    \label{eq:SSIM}
\end{equation}
where $\mu_s, \sigma_s, \sigma_{s\hat{s}}$ are the mean and variance of $\bm{S}$, and the covariance between $\bm{S}$ and $\hat{\bm{S}}$, respectively. $c_1$ and $c_2$ are constants for numeric stability.
{For both half- and full-duplex relays, we adopt the `bandwidth ratio' to quantify the available (complex) channel uses per pixel (CPP), defined as $\rho \triangleq \frac{k}{M}$.}

{\textbf{Remark}. For any given finite $k$, the block-based scheme presented above reduces to direct transmission for $B = 1$, and to half-duplex relay with $\alpha = 1/2$ for $B = 2$. In general, one would expect that its performance should increase with $B$ as the initial period in which the relay remains silent (for $k/B$ symbols) becomes shorter; however, this requires specifying the relay operation at every block; and therefore, results in increased complexity, making the design of a practical coding scheme infeasible.}

\section{DeepJSCC-based Half-Duplex Relaying}\label{sec:half-duplex}
In this section, we propose two protocols adopting DeepJSCC over the half-duplex relay channel, namely, the DeepJSCC-AF and DeepJSCC-PF. {In both protocols, we parameterize the encoder $f_{s, \Phi}(\cdot)$, decoder $g_{d, \Psi}(\cdot)$, and the transformation {$f_{r, \Theta}(\cdot)$} at the relay\footnote{{The transformation $f_{r, \Theta}(\cdot)$ used at the relay terminal depends on the particular relaying scheme employed. In the case of DeepJSCC-AF, the relay simply amplifies its received signal, and no DNNs are needed ($\Theta = \emptyset$).}} by DNNs, where $\Phi, \Theta, \Psi$ denote the neural network weights at the source, relay and the destination nodes, respectively.}


\subsection{Recap of ViT models}\label{sec:vit_enc}
Before delving into the details of the proposed relaying protocols, we provide a quick overview of the encoding and decoding processes using ViT models, which form the core foundation of our approach.

\subsubsection{ViT encoder}
As shown in {the left hand side of}  Fig. \ref{fig:vit_codec}, the ViT encoder is comprised of three parts: image-to-sequence transformation, self-attention, and linear projection.

\textbf{Image-to-sequence transformation.}  We evenly partition the input image $\bm{S}$ into a sequence of $p\times p$ tokens along its spatial dimensions\footnote{We assume both $H$ and $W$ are multiples of $p$, which can be ensured by zero-padding.}, where each token consists of {$N_t \triangleq {M}/{p^2}$} elements. The tokens are further processed by a multilayer perceptron (MLP) layer with Gaussian error linear unit (GeLU) activation function with an output of dimension $c$. 

\textbf{Self-attention module.} After obtaining the $p^2$ tokens, the same positional embedding technique in \cite{VIT} is adopted to provide additional positional information, and organize the positionally embedded tokens into a matrix $\bm{S}_e\in \mathbb{R}^{p^2\times c}$. Then, $N_e$ transformer layers are stacked together and applied to $\bm{S}_e$ to generate output $\widetilde{\bm{S}}_e$. As an example, we illustrate the operations of the first transformer layer as follows:
\begin{align}
    \bm{S}_1 &= \bm{S}_e + \text{MSA}(\bm{S}_e), \notag \\
    \bm{S}_2 &= \bm{S}_1 + \text{MLP}(\text{LN}(\bm{S}_1)),
    \label{eq:MSA}
\end{align}
where $\text{MSA}(\cdot)$ denotes the multi-head self-attention layer \cite{attention}, $\text{LN}(\cdot)$ represents layer norm operation and $\text{MLP}$ is comprised of linear layers with GeLU activation function. Note that $\bm{S}_2$ will be further fed into the subsequent transformer layers.

\textbf{Linear Projection.} After passing $N_e$ transformer layers, we apply a linear layer to the output matrix $\widetilde{\bm{S}}_e$ (or $\widetilde{\bm{S}}^\prime_e$ if LA module is adopted for adaptive transmission introduced in Section \ref{sec:adapt_trans}) with dimensions $p^2\times c$, and map it to the output matrix $\bm{X} \in \mathbb{R}^{p^2\times c^*}$, where $p^2 c^* = 2k$ and $k$ is the number of complex channel uses.

\subsubsection{ViT decoder}
\begin{figure}
\centering
\includegraphics[width=\linewidth]{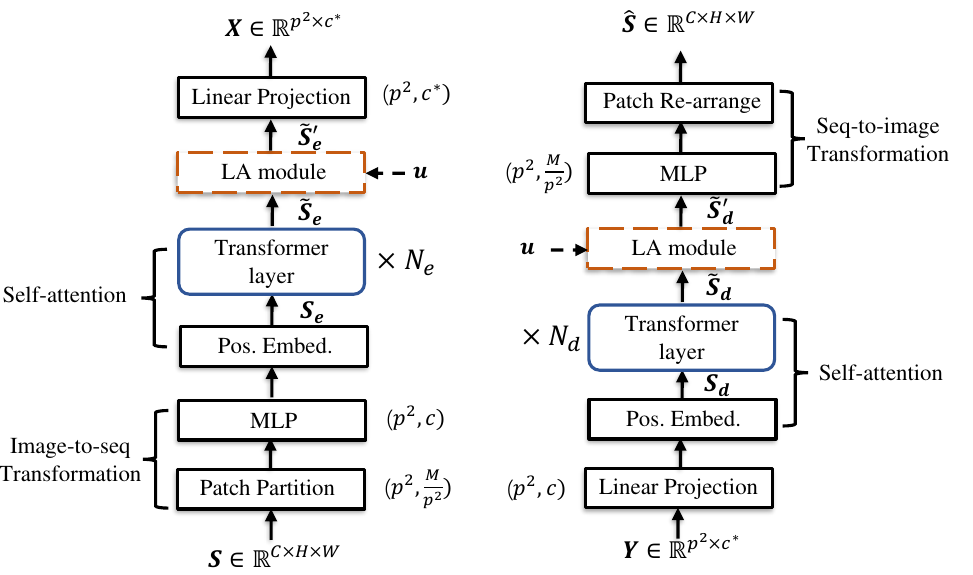}
\caption{The neural network architecture of the ViT encoder (left) and decoder (right). The (optional) `LA module' which takes the side information $\bm{u}$ as an additional input is adopted for the adaptive transmission model introduced in Section \ref{sec:fd_relay}.}
\label{fig:vit_codec}
\end{figure}

The ViT decoding process mirrors ViT encoding. As shown in {the right hand side of} Fig. \ref{fig:vit_codec}, the ViT decoder is also comprised of three modules: linear projection, self-attention, and sequence-to-image transformation. If we denote the noisy channel output by $\bm{Y} \in \mathbb{R}^{p^2 \times c^*}$, which is fed into the ViT decoder, the linear projection module maps each token of $\bm{Y}$ to a $c$-dimensional vector, which will be positionally embedded to form a matrix $\bm{S}_d\in \mathbb{R}^{p^2\times c}$ and further processed by the subsequent $N_d$ transformer layers. An MLP with GeLU activation function is applied to the final output of the self-attention modules $\widetilde{\bm{S}}_d$ ($\widetilde{\bm{S}}'_d$ if LA module is adopted) to generate a matrix with dimensions $p^2 \times \frac{M}{p^2}$, {where $M$ is the total number of pixels defined in \eqref{eq:psnr}}. Finally, the patch re-arrange layer is responsible for converting this matrix back to the reconstructed image $\hat{\bm{S}} \in \mathbb{R}^{C\times H\times W}$.

\subsection{Relaying Protocols}
This subsection introduces the proposed DeepJSCC-based relaying protocols. We will provide an in-depth exploration of the DNN structures tailored specifically to accommodate and execute these protocols seamlessly.

\subsubsection{DeepJSCC-AF}
We first consider a simple relaying scheme, where the relay transmits a scaled version of its received signal:
\begin{equation}
 \bm{x}_r = \eta \cdot \bm{y}_{r}, 
 \label{eq:af}
\end{equation}
where $\eta$ is chosen to satisfy the power constraint. Note that, in the DeepJSCC-AF protocol, to ensure the relay participates in the whole transmission process for better performance, we set $\alpha = 1/2$ such that the transmitted signal $\bm{x}_r\in \mathbb{C}^{(1-\alpha) k}$ is of the same length with that of $\bm{y}_r \in \mathbb{C}^{\alpha k}$. Then, $\eta$ can be calculated as $\eta =  \sqrt{\frac{2P_r}{P_s^{(1)} c_{sr}^2+1}}$, where $P_s^{(1)}$ denotes the source transmit power in the \textit{relay-receive} period.

At the destination, the received signal {consists of $\bm{y}_d^{(1)}$, received during} the \textit{relay-receive} period, $\bm{y}_d^{(1)}$, and {$\bm{y}_d^{(2)}$ received during} the \textit{relay-transmit} period.
The entire received signal, $\bm{y}_d$, at the receiver is converted into a real vector, reshaped to $\bm{Y} \in \mathbb{R}^{p^2\times c^*}$, and fed to the DeepJSCC-AF decoder parameterized by the aforementioned ViT decoder to generate the reconstructed image $\hat{\bm{S}}$. 

\subsubsection{DeepJSCC-PF}
DeepJSCC-AF suffers from noise propagation since the relay merely transmits a scaled version of the received noisy signal. In the realm of digital communications, more sophisticated relaying schemes, such as DF and partial DF (pDF) \cite{relay_capacity2}, address this issue.
In particular, the DF (pDF) protocol decodes the original message (or part of it) getting rid of the noise, and only relevant information is relayed to the destination. Denoising the received signal at the relay is challenging in the context of DeepJSCC-based schemes, which directly map the source information to continuous amplitude symbols. Inherently, complete denoising will not be possible in DeepJSCC, which renders both the DF and pDF schemes infeasible.

Another well-known relaying scheme, CF \cite{relay_capacity0, tit_cf}, involves the relay compressing its received signal treating the destination's received signal as correlated side information. Wyner-Ziv coding is used to compress the relay's received signal, whose index is forwarded to the destination using an independent channel code. In addition to the suboptimality of this separation-based approach in the finite block length regime, the compression scheme at the relay is also oblivious to the underlying channel code. In general, it is also possible to combine pDF and CF into the partial decode-compress-and-forward (pDCF) scheme \cite{orthogonal_relay}.

In an attempt to combine the benefits of the conventional pDF and CF protocols in the context of DeepJSCC, we introduce DeepJSCC-PF. In this scheme, {we parameterize the relay's processing function $f_{r, \Theta}(\cdot)$ by a modified ViT encoder}. It is worth noting that, unlike the source encoder, which receives an image as the input, there is no need for a patch partition layer in the modified ViT encoder at the relay, whose input is the received noisy channel output.
The underlying idea here is to craft a relay protocol that undergoes automatic optimization through end-to-end training, ultimately aiming for superior reconstruction performance. This approach allows us to break away from the limitations of conventional methods and adapt our relaying strategy to the specific needs of the source data, enhancing its overall effectiveness.

Following the intuition, the overall encoding, relay processing and decoding procedures can be summarized as follows: the ViT encoder at the source node generates $\bm{X} \in \mathbb{R}^{p^2\times c^*}$ as shown in Fig. \ref{fig:system_model}. Then, we partition it into two parts along the column\footnote{Note that it is also possible to divide the matrix $\bm{X}$ along the row direction. We find in experiments that both partition methods yield similar performance for both half-duplex and full-duplex cases, and we chose column partition to simplify our description. The comparable performance of the two methods is also reported in \cite{deepjscclpp} in the context of bandwidth-adaptive transmission.} for \textit{relay-receive} and \textit{relay-transmit} periods as $\bm{X} = [\bm{X}^{(1)}, \bm{X}^{(2)}]$, where each column of $\bm{X}^{(1)}$ and $\bm{X}^{(2)}$ is of $\alpha c^*$ and $(1-\alpha) c^*$ dimension, respectively.  The received signal at the relay, $\bm{y}_r$, is first converted to a real vector and reshaped into $\bm{Y}_r$ with dimensions $p^2 \times \alpha c^*$. A linear projection layer maps each token of $\bm{Y}_r$ to a $c$-dimensional vector followed by a positional embedding layer to form a new matrix that will be fed to $N_r$ consecutive transformer layers. Notice that the relay output is a complex vector of length-$(1-\alpha)k$, thus, the final linear projection layer will map each $c$-dimensional input token to a $(1-\alpha) c^*$-dimensional vector.

For both protocols, we adopt the mean square error (MSE) as the loss function {during training}: 
\begin{equation}
    \mathcal{L} = \mathbb{E}_{\bm{S}\sim p(\bm{S})}\left[\|\bm{S} - \hat{\bm{S}}\|^2_F\right].
    \label{equ:Loss_func}
\end{equation}

\subsection{Important variables} \label{sec:hd_variable}
For digital cooperative communications, when the link qualities $(c_{sr}, c_{rd}, c_{sd})$ and the transmission power $(P_s, P_r)$ are given, some variables, such as the time division variable $\alpha$, the power allocation variable $\gamma$, and the correlation variable $\beta$ (for DF protocol), are critical in determining the achievable rate. The conventional schemes choose these variables in order to maximize their achievable rate \cite{relay_capacity}. In this section, our intention is to highlight the important variables for the performance {of the DeepJSCC-PF scheme}. Later in the simulation results, we will demonstrate how the DNNs can autonomously learn to optimize these variables, ultimately leading to an efficient end-to-end image reconstruction process.

First, the time division variable $\alpha$ introduced in Section \ref{sec:system_model} divides the {available bandwidth between the} relay transmit and relay receive periods. Second, the power allocation variable $\gamma$ defined as below determines the average transmission power for the two periods, $P_s^{(1)}$ and $P_s^{(2)}$, defined as:
\begin{align}
    \gamma \triangleq \frac{\mathbb{E}{\|\bm{x}_s^{(1)}\|^2_2}}{\mathbb{E}{\|\bm{x}_s\|^2_2}},
    \label{eq:gamma_def}
\end{align}
which leads to $P_s^{(1)} = \frac{\gamma P_s}{\alpha}$ and $P_s^{(2)} = \frac{(1-\gamma) P_s}{1-\alpha}$.
Finally, we introduce the correlation variable $\beta$ as follows.
Consider the conventional pDF protocol where the source aims to convey an index $w \in [1, 2^{nR}]$ to the destination. {It first splits it into sub-indexes $w_1$ and $w_2$.} In the \textit{relay-receive} period, the source encodes $w_1$ to $\bm{x}^{(1)}_s(w_1)$ with power $P_s^{(1)}$, which is broadcasted to both the relay and the destination. The relay receives the message, decodes the index $w_1$ and re-encodes it to $\bm{x}_r = \bm{x}_r(w_1)$, which is power normalized to $P_r$. The source signal, $\bm{x}_s^{(2)} = \bm{x}_s^{(2)}(w_1, w_2)$ at the \textit{relay-transmit} period is comprised of two independent parts, $\bm{x}_r(w_1)$ and $\bm{x}_s^{(2)}(w_2)$, which are superimposed for transmission.  The first term is identical to the relay transmitted signal (up to a scaling factor) whose power is $\beta P_s^{(2)}$, while the second term conveys the distinct message $w_2$ and is independent of the first term with power $(1-\beta)P_s^{(2)}$. {Accordingly}, the correlation variable is given by 
\begin{equation}
    \beta = \frac{\mathbb{E}{(\bm{x}_r^\dagger \bm{x}_s^{(2)} })}{\sqrt{\mathbb{E}{\|\bm{x}_s^{(2)}\|^2_2} \mathbb{E}{\|\bm{x}_r\|^2_2}}}.
\end{equation}

Intuitively, the $\beta$ value indicates the `correlation' between the relay transmit signal and the source transmit signal in the \textit{relay-transmit} period. As long as the relay correctly decodes the index $w_1$, having a non-zero $\beta$ value allows the coherent superposition of the signals from the source and the relay to boost their power against the noise, providing stronger error correction ability for $w_1$. However, a large $\beta$ value implies the source {transmits only a limited amount of} new information $w_2$ to the destination resulting in a rate loss. Thus, it is important to figure out a good $\beta$ to balance the transmission between $w_1$ and $w_2$. {Even though we do not have the concepts of sub-messages in DeepJSCC-PF, $\beta$ value can still be used to quantify how much of the source power is used to collaborate with the relay, and how much of it for the transmission of new information.}

{Due to our end-to-end approach,} it is not possible to figure out the optimal parameters analytically as in the case of \cite{relay_capacity}, {instead, for each link quality and transmit power, we will first determine {the best $\alpha$ value, denoted by} $\alpha^*$, from a discrete set, and then let the neural network to automatically figure out the corresponding $\beta$ and $\gamma$ values for that $\alpha^*$.} As will be shown in the simulation part, by analyzing the variables $\alpha, \beta, \gamma$, {we can gain more insights on the behavior of the DeepJSCC-PF scheme.}

{The training algorithm of the DeepJSCC-PF protocol over half-duplex relay is summarized in Algorithm \ref{algorithm:hd_pf}.}

\begin{algorithm}[t]

	\caption{Training of DeepJSCC-PF over half-duplex relay.}
    \SetKwInOut{Input}{Input}
    \SetKwInOut{Output}{Output}
    \SetKwFunction{SymbolMapping}{SymbolMapping}
    \SetKwFunction{SymbolDeMapping}{SymbolDeMapping}
    \label{algorithm:hd_pf}
	
	\Input{$c_{sr}, c_{rd}, c_{sd}, P_s, P_r, \alpha, \{\bm{S}\}, N_{epoch}$.}
    \Output{Optimized $\{\Phi, \Theta, \Psi\}$.}
	
	\BlankLine
	   \For{$n=1$ \KwTo {$N_{epoch}$}}{
            \For{Each batch $\bm{S} \sim \{\bm{S}\}$}{
            $\bm{X} = f_{s, \Phi}(\bm{S})$; \\
            Partition $\bm{X} \triangleq [\bm{X}^{(1)}, \bm{X}^{(2)}]$, $\bm{X}^{(1)} \in \mathbb{R}^{p^2\times \alpha c^*}$; \\
            $\bm{x}_s^{(1)}, \bm{x}_s^{(2)} \leftarrow \SymbolMapping(\bm{X}^{(1)}, \bm{X}^{(2)})$;
            where $\|\bm{x}_s\|_2^2 \le kP_s$.
            \Comment{\textbf{Source encode.}} \\

            $\bm{y}_r = c_{sr}\bm{x}_s^{(1)} + \bm{n}_r$; \\
            $\bm{x}_r \leftarrow f_{r, \Theta}(\bm{y}_r)$; where $\|\bm{x}_r\|_2^2 \le kP_r$.\\ \Comment{\textbf{Relay encode.}} \\

            $\bm{y}_d^{(1)}, \bm{y}_d^{(2)} = c_{sd}\bm{x}_s^{(1)} + \bm{n}_d^{(1)}, c_{sd}\bm{x}_s^{(2)} + c_{rd}\bm{x}_r + \bm{n}_d^{(2)}$;\\
            $\bm{Y}_d \leftarrow \SymbolDeMapping([\bm{y}_d^{(1)}, \bm{y}_d^{(2)}])$; \\
            $\hat{\bm{S}} = g_{d, \Psi}(\bm{Y}_d)$; \Comment{\textbf{Destination decode.}}
        
            Optimize parameters $\{\Phi, \Theta, \Psi\}$ via gradient descent using $\mathcal{L} = \|\bm{S} - \hat{\bm{S}}\|_F^2$.
            }}

\end{algorithm}

\section{DeepJSCC-based Full-Duplex Relaying}\label{sec:fd_relay}
In this section, we delve into the realm of full-duplex relaying. We shall elucidate the means by which enhanced performance can be realized when the relay possesses the capability to receive and transmit simultaneously.

{In conventional full-duplex relaying schemes, DF and CF, which focus on channel coding, block coding can be trivially employed by dividing the message into $B$ independent submessages. The relay forwards the message/signal it receives in channel block $b$ from the source to the destination in block $b+1$. The source transmits a new submessage in each channel block, but can also use part of its power to collaborate with the relay in forwarding the previous submessage (as in the DF scheme for full-duplex relays). To guarantee that each submessage can be transmitted at the same rate in cooperation with the relay, $B$ submessages are transmitted over $B+1$ channel blocks, where the relay remains silent in the first channel block and the source does not transmit a new submessage in the last one.

In the asymptotic regime of $k \rightarrow \infty$, we can have both $B \rightarrow \infty$ and $k/B \rightarrow \infty$, which means that each submessage can be decoded reliably, and sending $B$ submessages over $B+1$ channel blocks results in a negligible rate loss. However, in practice (i.e., finite $k$ and $B$), there is a trade-off between the number ($B$) and length ($k/B$) of the blocks. We want large $B$ to minimize rate loss, but this results in reduction in the block length of individual submessages, resulting in increased error probability.
}

\subsection{DeepJSCC with Block Transmission}
{Similarly to the half-duplex relay channel, we can consider both DeepJSCC-AF and DeepJSCC-PF for the full-duplex case.}

\subsubsection{DeepJSCC-PF}
We note that DF with BMC cannot be applied directly to DeepJSCC since perfect decoding is not possible. Instead, we introduce a novel block-based DeepJSCC scheme, where the source transmits the input image in multiple blocks {as described in Section \ref{sec:sec_IIB}. However, unlike in channel coding, we cannot easily divide our input into $B$ independent parts with equal information content. In the case of JSCC, this would require $B$ statistically equivalent parts so that we can use the same encoding and decoding functions for each of them. Instead, the source maps the input to one long channel codeword of $k$ symbols. This means that the signal forwarded by the relay during each channel block depends on the signal it has received from the source over all the previous blocks; and hence, the relay encoder will acquire more and more information as time goes by. We expect the relay to gradually refine its estimate of the input image, and transmit increasingly relevant information as time progresses. Therefore, in principle, we need a separate relay encoding function for each channel block; however, training a separate encoder for each block will be computationally prohibitive, especially when the number of blocks, $B$ is large.}

\textbf{Source encoding.} 
{The encoding function $f_{s, \Phi}(\cdot)$, parameterized by a ViT encoder with parameters, $\Phi$}, directly maps the source $\bm{S}$ to $\bm{x}_{s}$ which will be further partitioned into $B$ blocks, $\bm{x}_s = [\bm{x}_{s,1}^\top,\ldots, \bm{x}_{s, B}^\top]^\top$.

\begin{table}[t]

\centering
\caption{{The block transmission used by the proposed DeepJSCC-PF protocol over the full-duplex relay channel.}}
\newcommand{\tabincell}[2]{\begin{tabular}{@{}#1@{}}#2\end{tabular}}
\begin{tabular}{cccccc}
\hline
$b$ &  1  &  2 &  \ldots & $B-1$ & $B$ \\ \hline
$\mathrm{S}$ & $\bm{x}_{s,1}$ & $\bm{x}_{s,2}$ & \ldots & $\bm{x}_{s,B-1}$ & $\bm{x}_{s,B}$\\ \hline
$\mathrm{R}$  & - & $f^2_{r,\bm{\Theta}}(\bm{x}_{s,1})$ & \ldots & \tabincell{c}{$f^{B-1}_{r,\bm{\Theta}}(\bm{x}_{s,1:B-2},$ \\ $ \bm{x}_{r,1:B-2})$} & \tabincell{c}{$f^B_{r,\bm{\Theta}}(\bm{x}_{s,1:B-1},$ \\ $ \bm{x}_{r,1:B-1})$}  \\
\hline
\label{my_schedule}
\end{tabular}

\end{table}

\begin{figure*}
\centering
\includegraphics[width=0.9\linewidth]{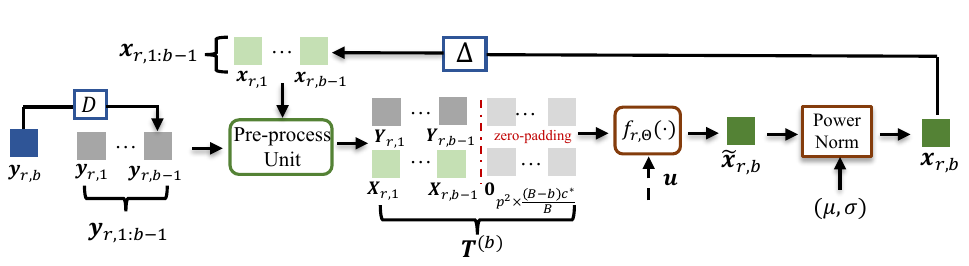}
\caption{The signal generation process of the $b$-th block at the relay node in the full-duplex relay case. $\Delta$ denotes unit delay. The `Pre-process Unit' first converts the vectors $\bm{x}_{r,j}, \bm{y}_{r,j}$ to their corresponding matrices, $\bm{X}_{r,j}, \bm{Y}_{r,j}, j\in [1, b-1]$, and then performs zero padding to obtain $\bm{T}^{(b)}$. The variables $\mu$ and $\sigma$ are introduced in \eqref{eq:mean_var} for power normalization.}
\label{fig:relay_flow}
\end{figure*}

\textbf{Relay encoding.}
At the $b$-th block, the relay has access to all the past received signals, $\bm{y}_{r,1:(b-1)}$, and its own transmitted signals, $\bm{x}_{r,1:(b-1)}$. All these signals will be given as input to the relay encoding function {$f^b_{r, \bm{\Theta}}(\cdot)$, which is a DNN parameterized by $\bm{\Theta}$.}  Taking all the (available) signals as input at the relay will improve the final performance, which is verified via numerical experiments in Section \ref{sec:experiment}. The relay transmit signal at the $b$-th block can be expressed as:
{\begin{equation}
    \bm{x}_{r,b} = f^b_{r,\Theta}(\bm{y}_{r,1:(b-1)}, \bm{x}_{r,1:(b-1)}).
\end{equation}}
We note here that the term $\bm{x}_{r,1:(b-1)}$ is equivalent to  $\bm{x}_{r,2:(b-1)}$ as the relay keeps silent in the first block, yet we will adopt the former notation for consistency. The transmitted signals of the source and relay at different blocks for DeepJSCC-PF are summarized in Table \ref{my_schedule}.

{We remark again that, unlike in channel coding, in JSCC over the relay channel, we cannot employ the exact same code at each channel block $b$ to transmit the corresponding submessage. Instead, we need to design a separate relay function with a different input size and statistics. As shown in Sec. \ref{sec:dnn_design}, we will present a unique relay encoder architecture which will be trained to be deployed at each channel block, significantly reducing the complexity.}

\textbf{Destination Decoding.}
As opposed to backward decoding employed in BMC, where the destination decodes the messages in a reverse order, our DeepJSCC decoder, {denoted by $g_{d, \Psi}(\cdot)$,} takes all the $B$ received blocks together to reconstruct the original image $\hat{\bm{S}}$, which can be expressed as:
\begin{equation}
    \hat{\bm{S}} = g_{d, \Psi}(\bm{y}_{d, 1:B}).
\end{equation}

\subsubsection{DeepJSCC-AF}
The operations at the source and the destination nodes of the DeepJSCC-AF protocol are identical to those of DeepJSCC-PF. The only difference lies in the transmitted signal at the relay, where the $b$-th block can be expressed as:
\begin{equation}
    \bm{x}_{r, b} = \eta \bm{y}_{r,b-1},
\end{equation}
where $\eta$ is the power normalization factor. Both DeepJSCC-AF and DeepJSCC-PF protocols for the full-duplex relay channel use the MSE between $\bm{S}$ and $\hat{\bm{S}}$ as the loss function as in the case of half-duplex relaying.

\subsection{DNN design}\label{sec:dnn_design}
Next, we investigate the intricacies of designing the DNNs at the source, relay, and destination to effectively accommodate the full-duplex relaying protocol introduced above.

\subsubsection{DNNs for the encoder}
Similarly to the half-duplex case, {the encoding function $f_{s, \Phi}(\cdot)$} at the source is realized using a standard ViT encoder, which takes the image as input and outputs a matrix $\bm{X}_s\in \mathbb{R}^{p^2\times c^*}$. Then, $B$ sub-matrices are obtained by partitioning along the columns expressed as:
\begin{equation}
    \bm{X}_s = [\bm{X}_{s, 1},\ldots, \bm{X}_{s, B}],
\end{equation}
where each $\bm{X}_{s,b} \in \mathbb{R}^{p^2\times \frac{c^*}{B}}$ is converted into a complex vector of length $k/B$ for transmission. 

\subsubsection{DNN at the relay}
The relay adopts a modified ViT encoder as its backbone. {To avoid training separate relay encoders for different channel blocks, we construct a sequence-to-sequence encoder built upon the transformer architecture. In this architecture, the relay encoder receives as input a sequence, $\bm{T}^{(b)}$, called the `knowledge matrix', which consists of the channel input and output signals at the relay up to that time, augmented with $0$ entries for the future channel inputs and outputs. This is fed into the transformer architecture, which, thanks to the positional encoder, learns to encode the available information to the appropriate channel input at each channel block.}

Since its input is no longer an image but a `\textit{knowledge matrix}' $\bm{T}\in \mathbb{R}^{p^2\times \frac{2c^*(B-1)}{B}}$, consisting of all the previous received signals as well as its own past transmitted signals, the patch partition operation is removed from the standard ViT encoder at the relay. The construction of the knowledge matrix $\bm{T}$, which corresponds to the `Pre-process Unit' in Fig. \ref{fig:relay_flow}, is described as follows: At the first block, $\bm{T}^{(1)}$ is initialized as an all-zero matrix as the relay has not yet received any signal. At the end of the $(b-1)$-th block, the relay has received $(b-1)$ blocks, $\bm{y}_{r,1:(b-1)}$, from the source and each block, $\bm{y}_{r,j}$ is a length-$k/B$ complex vector converted to a matrix $\bm{Y}_{r, j} \in \mathbb{R}^{p^2\times \frac{c^*}{B}}, j \in [1, b-1]$. Similarly, each transmitted signal $\bm{x}_{r, j}$ is organized into $\bm{X}_{r, j} \in \mathbb{R}^{p^2\times \frac{c^*}{B}}$ and the updated knowledge matrix at the relay is simply:
\begin{align}
    \bm{T}^{(b)} &= [\bm{Y}_{r, 1}, \ldots, \bm{Y}_{r, b-1}, \bm{0}_{p^2\times  \frac{(B-b)c^*}{B}}, \notag \\ 
    &\bm{X}_{r, 1}, \ldots, \bm{X}_{r, b-1}, \bm{0}_{p^2\times \frac{(B-b)c^*}{B}}],
    \label{eq:update_tensor_T}
\end{align}
where $\bm{0}_{p^2\times \frac{(B-b)c^*}{B}}$ denotes an all-zero matrix with dimension $(p^2,\frac{(B-b)c^*}{B})$. In the $b$-th block, the relay encoder $f_{r, \bm{\Theta}}(\cdot)$ takes $\bm{T}^{(b)}$ as input and outputs $\tilde{\bm{x}}_{r,b} \in \mathbb{R}^{\frac{2k}{B}}$:
\begin{equation}
    \tilde{\bm{x}}_{r,b} = f_{r, \bm{\Theta}}(\bm{T}^{(b)}).
\end{equation}
We remark that the relay starts from an all-zero knowledge matrix $\bm{T}^{(1)}$, and gradually populates it as blocks progress, and we expect the ViT encoder at the relay to acquire increasingly higher quality features about the source image to be forwarded to the destination.

\textbf{Power normalization at the relay.} Before transmitting $\tilde{\bm{x}}_{r,b}$ to the destination, a power normalization layer is needed to satisfy the relay power constraint. At a first glance, the same power normalization at the source can be adopted at the relay node. However, since the relay has no access to the future blocks with index $j>b$ when performing power normalization for the current block with index $b$, a new power normalization scheme is needed. To tackle this, we adopt the scheme used in the deep learning-aided feedback code design\cite{gbaf}. During training, we record the mean $\mu$ and variance $\sigma^2$ of $\tilde{\bm{x}}_r$ of the transmitted symbols:
\begin{align}
    \mu &= \frac{B}{2k(B-1)}\sum_{b=2}^B \sum_{n=1}^{2k/B} {\tilde{\bm{x}}_{r, b}[n]}, \notag \\
    \sigma^2 = &\frac{B}{2k(B-1)} \sum_{b=2}^B \sum_{n=1}^{2k/B} {(\tilde{\bm{x}}_{r, b}[n] - \mu)^2}.
    \label{eq:mean_var}
\end{align}
Note that the calculation of $\mu, \sigma$ starts from block $b = 2$ since the relay keeps silent in the first block.
At the inference time, the power normalization is performed in a block wise manner:
\begin{equation}
    \tilde{\bm{x}}^\prime_{r,b} = \sqrt{\frac{Pr}{2}} \frac{\tilde{\bm{x}}_{r,b} - \mu}{\sigma}, \quad  b \in [2, B]
    \label{eq:power_norm_fd}
\end{equation}
where $\sqrt{1/2}$ is introduced due to the fact that $\tilde{\bm{x}}_{r,b}$ is a real-valued vector. Finally, we convert $\tilde{\bm{x}}^\prime_{r,b}$ to a complex vector, $\bm{x}_{r,b} \in \mathbb{C}^{\frac{k}{B}}$, which will be forwarded to the destination. {Fig. \ref{fig:relay_flow} illustrates the entire relay encoding process for the $b$-th block.}


\subsubsection{DNNs for the decoder}
Finally, the decoder at the destination node, {denoted as $g_{d, \Psi}(\cdot)$,} takes all the received signals as input to reconstruct the original image $\bm{S}$. Similarly to the processing at the relay, the received signal at the destination at the $b$-th block, $\bm{y}_{d, b}$ is converted into a matrix $\bm{Y}_{d, b} \in \mathbb{R}^{p^2\times \frac{c^*}{B}}$ and we concatenate all the $B$ matrices along the columns, $\bm{Y}_{d} = [\bm{Y}_{d, 1}, \ldots, \bm{Y}_{d, B}]$, which will be fed to the {decoder function, $g_{d, \Psi}(\cdot)$.} A standard ViT decoder model introduced in Section \ref{sec:vit_enc} with $N_d$ attention blocks is utilized to parameterize {$g_{d, \Psi}(\cdot)$.}

\begin{figure}
\centering
\includegraphics[width=0.9\linewidth]{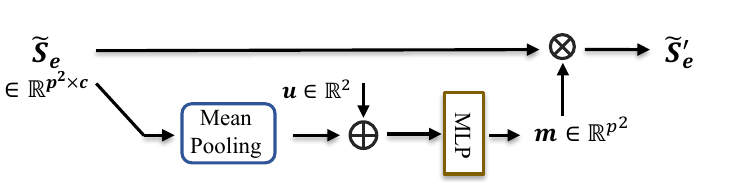}
\caption{The neural processing at the LA model located at the ViT encoder, where $\oplus$ denotes vector concatenation, and $\otimes$ denotes token-wise multiplication.}
\label{fig:LA_module}
\end{figure}

\subsection{Adaptive Transmission}\label{sec:adapt_trans}
As shown in Fig. \ref{fig:system_model:b}, the full-duplex relay channel is determined by variables such as link qualities $c_{sr}, c_{sd}, c_{rd}$ and the average transmission power at the source and the relay, i.e., $P_s$ and $P_r$, respectively.
If a new model is trained for each and every tuple of these variables, the memory cost for storing these models would be prohibitive. Thus, to make the DeepJSCC protocols more practical, training a single model to be adaptive to different variables is desired. Note that \cite{xu2021wireless, deepjscclpp} have shown that SNR-adaptive transmission can be achieved over the AWGN channel by introducing self-attention module. {However, it is not clear if the self-attention module is still effective in the cooperative relay settings}.

{To begin with, we simplify the problem by setting $c_{sd}$ to unity while $c_{sr}, c_{rd} \in [c_{min}, c_{max}]$ dB and $P_s, P_r \in [P_{min}, P_{max}]$ dB. }
To train the adaptive model, we first reveal the side information, denoted as {$\bm{u} \triangleq [c_{sr}, c_{rd}, P_s, P_r]$}, to the source, relay and destination nodes. A new neural network architecture, termed as link-adaptive (LA) module at these nodes utilizes the side information to facilitate the generation of the output signal for the specific {link qualities and transmission powers}. {We illustrate the LA module at the source node as an example while the processing of the LA modules at the relay (if DeepJSCC-PF is adopted) and destination are the same.} As shown in Fig. \ref{fig:vit_codec}, the LA module is placed directly after the self-attention modules. Denoting the output of the self-attention module as $\widetilde{\bm{S}}_e \in \mathbb{R}^{p^2 \times c}$, the LA module takes $\widetilde{\bm{S}}_e$ and the side information as input and learns to assign different weights $\bm{m}$ to different tokens belonging to $\widetilde{\bm{S}}_e$:
\begin{align}
    \bm{m} &=  
    \text{MLP}\left(\left[\frac{1}{c}\sum_{j=1}^{c}\widetilde{\bm{S}}_e[1,j],  \ldots, \frac{1}{c}\sum_{j=1}^{c}\widetilde{\bm{S}}_e[p^2,j],\bm{u}\right]\right), \notag \\
    {\widetilde{\bm{S}}_e^\prime} & = \bm{m} \otimes \widetilde{\bm{S}}_e,
    \label{eq:modulated}
\end{align}
where $\bm{m}$ is of length-$p^2$ and $\otimes$ represents token-wise multiplication. Flowchart of the LA module is shown in Fig. \ref{fig:LA_module}. 

In the training phase, for each batch, {we randomly sample $(c_{sr}, c_{rd}) \sim \mathcal{U}(c_{min}, c_{max})$, $(P_s, P_r)  \sim \mathcal{U}(P_{min}, P_{max})$}, and feed them to the DNNs at the source, relay and destination node, respectively. All these DNN models are jointly optimized via end-to-end training. We will show in the simulation section that with the adaptive transmission model, our scheme can achieve comparable reconstruction performance with respect to the separately trained models for {each tuple of $(c_{sr}, c_{rd}, P_s, P_r)$}.

{Finally, we summarize the overall training process of the DeepJSCC-PF protocol over the full-duplex relay with adaptive transmission in Algorithm \ref{algorithm:fd_pf}.}

{
\subsection{Fading Channel} \label{sec:fading}
Finally, we consider the wireless fading channel with both large-scale and small-scale fading effects for the full-duplex relay\footnote{Our proposed scheme is also applicable to the half-duplex relay yet we omit the detailed description due to the page limit.}. The input-output relation of the Rayleigh fading channel can be expressed as:
\begin{equation}
    \bm{y} = h d^{-\tau} \bm{x} + \bm{n},
\end{equation}
where $h \sim \mathcal{CN}(0, 1)$ denotes the small-scale fading coefficient, $d$ is the distance between the transmitter and the receiver, $\tau$ is the attenuation  coefficient, and $\bm{n} \in \mathcal{CN}(\bm{0}, \sigma^2\bm{I})$ is the noise term. Note that the link qualities, $c_{sr}, c_{rd}, c_{sd}$ have already modeled the large-scale fading. Thus, in this subsection, we additionally consider the small-scale fading for the three links and denote the small-scale  fading coefficients for the S-R, R-D and S-D links as $h_{sr}, h_{rd}$ and $h_{sd}$, respectively. We further assume that all the three nodes, i.e., the source,  relay and the destination node have access to all the CSI information, i.e, $h_{sr}, h_{rd}, h_{sd}$ and $c_{sr}, c_{rd}, c_{sd}$. With the CSI, the transmitter of the source and the relay nodes perform precoding operations for a better performance which is detailed as follows.

Since the processing for the DeepJSCC-PF and DeepJSCC-AF are similar, we only illustrate the former in the full-duplex relay mode. In particular, the source node precodes its encoded signal, $\tilde{\bm{x}}_{s, b}$ as:
\begin{equation}
    \bm{x}_{s, b} = \frac{h_{sd}^*}{|h_{sd}|} \tilde{\bm{x}}_{s, b},
\end{equation}
where $\tilde{\bm{x}}_{s, b}$ corresponds to the power normalized $b$-th transmission block. The relay received signal can be expressed as:
\begin{equation}
    \bm{y}_{r, b} = h_{sr}c_{sr}\frac{h_{sd}^*}{|h_{sd}|} \tilde{\bm{x}}_{s, b} + \bm{n}_{r,b},
\end{equation}
where $\bm{n}_{r, b} \in \mathcal{CN}(\bm{0}, \sigma_r^2\bm{I})$.
Since the relay knows $h_{sr}, h_{sd}$ and $c_{sr}$, it equalizes the received signal, $\bm{y}_{r, b}$ as:
\begin{equation}
    \hat{\bm{x}}_{s, b} = \frac{h_{eff}^*}{\sqrt{h_{eff}^*h_{eff} + \sigma_r^2/P_s}} \bm{y}_{r, b},
\end{equation}
where $h_{eff} \triangleq \frac{h_{sr}c_{sr}h_{sd}^*}{|h_{sd}|}$. Then, the relay node encodes the power normalized equalized signal to generate $\bm{\tilde{x}}_{r, b}$  which is precoded before transmission as:
\begin{equation}
    \bm{x}_{r, b} = \frac{h_{rd}^*}{|h_{rd}|} \tilde{\bm{x}}_{r, b}.
\end{equation}
The received signal at the destination node can be expressed as:
\begin{align}
    \bm{y}_{d, b} &= h_{sd}c_{sd}\bm{x}_{s, b} + h_{rd}c_{rd}\bm{x}_{r, b} + \bm{n}_{d, b}\notag \\
    &= |h_{sd}|c_{sd}\bm{\tilde{x}}_{s, b} + |h_{rd}|c_{rd} \bm{\tilde{x}}_{r, b} + \bm{n}_{d, b}.
\end{align}
We emphasize that, thanks to the precoding operations at the source and the relay, the phase term, $\phi \triangleq \text{arg}(h)$ is eliminated for a better decoding performance at the receiver.}

\begin{algorithm}[t]

    \caption{Training algorithm for the DeepJSCC-PF protocol over full-duplex relay with adaptive transmission.}
    \SetKwInOut{Input}{Input}
    \SetKwInOut{Output}{Output}
    \SetKwFunction{SymbolMapping}{SymbolMapping}
    \SetKwFunction{SymbolDeMapping}{SymbolDeMapping}
    \SetKwFunction{UpdateKnowledgeTensor}{UpdateKnowledgeTensor}
    \label{algorithm:fd_pf}
	\Input{$c_{min}, c_{max}, P_{min}, P_{max}, B, N_{epoch}, \{\bm{S}\}$.}
	\Output{Optimized $\{\Phi, \Theta, \Psi\}$.}
	\BlankLine
        
	\For{$n=1$ \KwTo {$N_{epoch}$}}{
        \For{Each batch $\bm{S} \sim \{\bm{S}\}$}{
            Sample  $c_{sr}, c_{rd} \sim \mathcal{U}(c_{min}, c_{max})$, $P_s, P_r \sim \mathcal{U}(P_{min}, P_{max})$; \\
            $\bm{u} = [c_{sr}, c_{rd}, P_s, P_r]$; \\
            $\bm{X} = f_{s, \Phi}(\bm{S}, \bm{u})$; \\
            Equally partition $\bm{X} \triangleq [\bm{X}_{s, 1}, \ldots, \bm{X}_{s, B}]$; \\
            $[\bm{x}_{s, 1}, \ldots, \bm{x}_{s, B}] \leftarrow \SymbolMapping([\bm{X}_{s, 1}, \ldots, \bm{X}_{s, B}])$; where $\|\bm{x}_s\|_2^2 \le kP_s$.
             \Comment{\textbf{Source encode.}} \\

            $\bm{T}^{(1)} \leftarrow \bm{0}$; \\
            
            \For{$b=1$ \KwTo {$B$}}{
                $\bm{y}_{r, b} = c_{sr}\bm{x}_{s, b} + \bm{n}_{r, b}$;\\
                $\tilde{\bm{x}}_{r,b} = f_{r, \bm{\Theta}}(\bm{T}^{(b)}, \bm{u})$; \\
                
                $\bm{T}^{(b+1)} \leftarrow \UpdateKnowledgeTensor(\bm{T}^{(b)}, \bm{y}_{r, b}, \tilde{\bm{x}}_{r,b})$ according to \eqref{eq:update_tensor_T}; \\
                
            }
            Calculate $\mu, \sigma^2$ based on $\tilde{\bm{x}}_{r,b}, b\in [2, B]$ according to \eqref{eq:mean_var}; \\
            Set $\tilde{\bm{x}}^\prime_{r,1} = \bm{0}$; Power normalize $\tilde{\bm{x}}^\prime_{r,b} = \sqrt{\frac{Pr}{2}} \frac{\tilde{\bm{x}}_{r,b} - \mu}{\sigma}, \; b\in [2, B];$ \\
            ${\bm{x}}_{r,b} \leftarrow \SymbolMapping(\tilde{\bm{x}}^\prime_{r,b})$; \Comment{\textbf{Relay encode.}} \\
                
            $\bm{y}_{d, b} = c_{sd}\bm{x}_{s, b} + c_{rd}\bm{x}_{r, b} + \bm{n}_{d,b}, \; b\in [1, B]$;\\
            $\bm{Y}_d \leftarrow \SymbolDeMapping([\bm{y}_{d, 1}, \ldots, \bm{y}_{d, B}])$; \\
            $\hat{\bm{S}} = g_{d, \Psi}(\bm{Y}_d, \bm{u})$; \Comment{\textbf{Destination decode.}}\\
        
            Optimize parameters $\{\Phi, \Theta, \Psi \}$ via gradient descent using $ \mathcal{L} = \|\bm{S} - \hat{\bm{S}}\|_F^2$. \\
            }}

\end{algorithm}


\newcolumntype{Y}{>{\centering\arraybackslash}X}

\begin{table}[t]

\caption{{List of key variables}}
\centering
 \begin{tabularx}{\linewidth}{| p{1cm} | Y |} 
 \hline

 Variable & Description \\  \hline

 $c_{sd}, h_{sd}$ & Channel coefficients for the source-to-destination link, set to unity by default.  \\  \hline
 $c_{sr}, h_{sr}$ & Channel coefficients for the source-to-relay link.  \\  \hline
 $c_{rd}, h_{rd}$ & Channel coefficients for the relay-to-destination link.\\ \hline
 {$P_s$} & Average transmission power of the source node.  \\  \hline
 {$P_r$} & Average transmission power of the relay node.  \\  \hline
 {$R^*_{hd}$} & Rate achieved by the maximum of the DF and CF protocols in the half-duplex mode.  \\  \hline
 {$R^*_{fd}$} & Rate achieved by the maximum of the DF and CF protocols in the full-duplex mode.  \\  \hline

 $\alpha$ & Time-division variable defined as the proportion of the \textit{relay-transmit} period over the entire period. \\  \hline 
 $\gamma$ & Power allocated to $\bm{x}_s^{(1)}$ at the source node in the half-duplex mode, defined as $\gamma = \frac{\mathbb{E}{\|\bm{x}_s^{(1)}\|^2_2}}{\mathbb{E}{\|\bm{x}_s\|^2_2}}$. \\  \hline 
 $\beta$ & Correlation coefficient between $\bm{x}^{(2)}_s$ and $\bm{x}_r$ in the half-duplex mode, defined as $\beta = \frac{\mathbb{E}{(\bm{x}_r^\dagger \bm{x}_s^{(2)} })}{\sqrt{\mathbb{E}{\|\bm{x}_s^{(2)}\|^2_2} \mathbb{E}{\|\bm{x}_r\|^2_2}}}$. \\  \hline 
 $B$ & Number of blocks in the full-duplex mode.\\  \hline 
 $t$ & Number of previous blocks the relay encoder takes as input in the full-duplex mode. \\  \hline 
 $p$ & The number of partitions along the height and width of the image, satisfying $p^2 = {M}/{N_t}$. \\  \hline 
 $c$ & Dimension of the hidden MLP layers in the ViT model. \\  \hline 
 \end{tabularx}
\label{tab:list_variables}

\end{table}

\section{Numerical Experiments}\label{sec:experiment}

\subsection{Parameter Settings and Training Details}
We evaluate the effectiveness of the proposed DeepJSCC-AF and PF protocols in both half-duplex and full-duplex relay channels considering the transmission of images from the CIFAR-10 dataset, which consists of $50,000$ training and $10,000$ test RGB images with $32 \times 32$ resolution. 

The ViT modules are used as the backbone for the DeepJSCC models. In particular, we set the parameter $p$ for the image-to-sequence transformation module to $8$, the number of hidden neurons in the MLP layers to $c = 256$, and unless otherwise mentioned, the number of transformer layers at the source, relay (if DeepJSCC-PF is adopted) and destination nodes are set to 6, 4, and 8, respectively.

In the training phase, Adam optimizer is adopted with a varying learning rate, initialized to $10^{-4}$ and reduced by a factor of $0.9$ if the validation loss does not improve in $20$ consecutive training epochs. The batch size for training is $64$ and the maximum number of epochs to train the models is set to $2\times 10^3$ to ensure that the reconstruction performance saturates with respect to the number of epochs. To avoid potential waste of computing resources during training, early stopping is used where the training process terminates if the validation loss does not improve over $60$ epochs. {For clarity, we present the important variables in Table \ref{tab:list_variables}.}

Unless otherwise mentioned, for both half-duplex and full-duplex models, we fix a CPP of $\rho = 0.25$, and hence, $c^* = 24$. We assume that the relay lies in between the source and the destination with $c_{sr}, c_{rd} \in [0, 10]$ dB while $c_{sd} = 0$ dB. {Moreover, simplifications are made when identical average power is considered at the source and the relay, where we denote $P_s = P_r \triangleq P$.} {Finally, we provide the $95\%$ confidence interval of the PSNR and SSIM performances for both the proposed schemes and the BPG baseline.}

\begin{figure}
\centering
\includegraphics[width=0.75\linewidth]{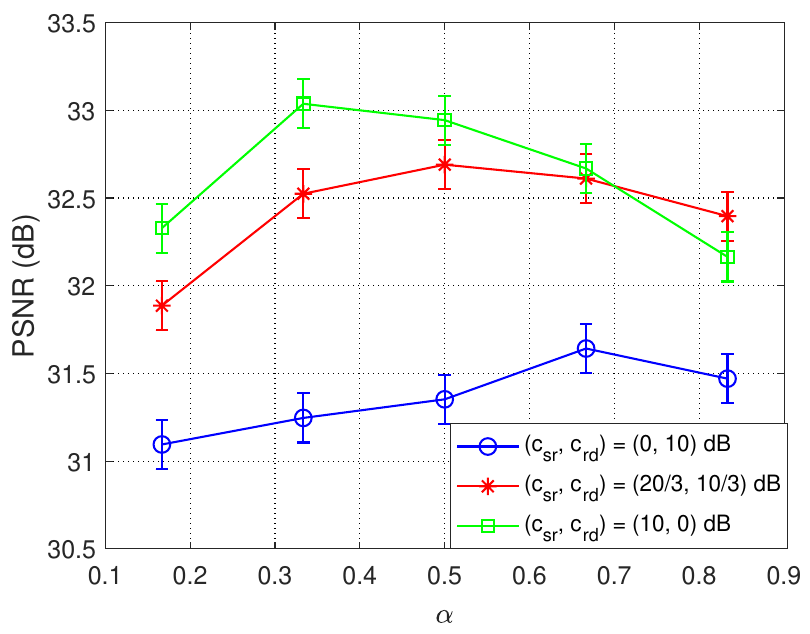}
\caption{{The PSNR performance for the half-duplex relay channel with different $\alpha$ values, $\alpha \in \{1/6, 2/6, 3/6, 4/6, 5/6\}$. The system settings are $(c_{sr}, c_{rd}) = (0, 10)$ dB, $(c_{sr}, c_{rd}) = (20/3, 10/3)$ dB and $(c_{sr}, c_{rd}) = (10, 0)$ dB, respectively.}}
\label{fig:hd_alpha}
\end{figure}

\begin{figure*}
\centering
\includegraphics[width=0.9\linewidth]{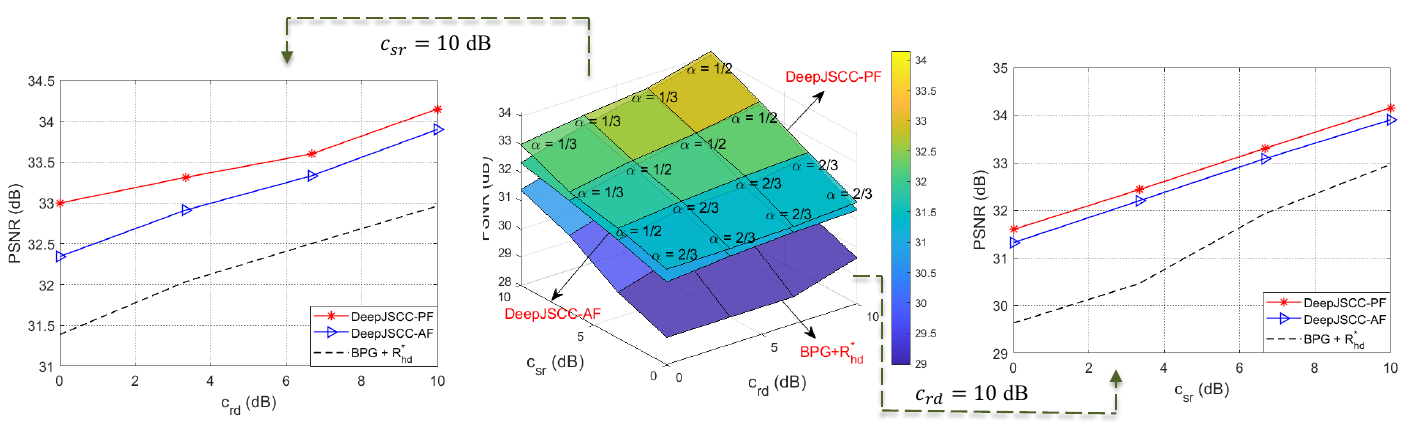}
\caption{{The PSNR performance of the proposed DeepJSCC-AF, DeepJSCC-PF and the digital baseline for the half-duplex relay channel with $c_{sr}, c_{rd} \in \{0, 10/3, 20/3 , 10\}$ dB.}}
\label{fig:hd_pf_vs_bpg}
\end{figure*}

\subsection{Performance evaluation for half-duplex relay}

\subsubsection{Overall DeepJSCC performance}
We first evaluate the relative performance of the proposed DeepJSCC protocols over the state-of-the-art BPG compression algorithm delivered at a rate\footnote{Since the general relay capacity is unknown, we provide the baseline results with two well-known relaying protocols.}, $R^*_{hd}$, achieved by the maximum of the conventional decode-and-forward ($R_{hd}^d$) and compress-and-forward ($R_{hd}^c$) protocols \cite{relay_capacity0,relay_capacity} working in a half-duplex mode. To be precise, we have:
\begin{align}
    R^*_{hd} = \max{(R_{hd}^d, R_{hd}^c)}.
    \label{eq:r_star}
\end{align}
{The closed form expressions for $R_{hd}^d$ and $R_{hd}^c$ and their dependence on the variables $P_s, P_r, c_{sr}, c_{rd}, c_{sd}$, and the time division variable $\alpha$ are provided in \cite{relay_capacity}. We also note that $R_{hd}^d$ and $R_{hd}^c$ are achieved using BMC and Wyner-Ziv coding \cite{nit} scheme in the literature.}

We first show numerically that for the DeepJSCC-PF protocol, there exists an optimal $\alpha$ for each channel condition. In this simulation, we consider three relay channel conditions, namely, (1) $(c_{sr} = 0, c_{rd} = 10)$ dB; (2) $(c_{sr} = 20/3, c_{rd} = 10/3)$ dB and (3) $(c_{sr} = 10, c_{rd} = 0)$ dB and for each of the scenarios, $\alpha$ is selected from a discrete set, $\{\alpha\} = \{1/6, 2/6, \ldots, 5/6\}$. The relative performance of the three settings with respect to different $\alpha$'s is shown in Fig. \ref{fig:hd_alpha}. {Note that the vertical lines correspond to the $95\%$ confidence intervals of the PSNR performance.} {As can be seen, the system prefers a smaller $\alpha$ when $c_{sr}$ is relatively large, while a larger $\alpha$ is more beneficial if $c_{rd}$ becomes larger. This observation is aligned with the intuition, when the source-to-relay channel is strong, relay can receive the necessary information over a shorter time period.} Note that this observation aligns with that of the conventional decode-and-forward protocol \cite{relay_capacity}, where the higher achievable rate, $R_{\text{DF}}$, is obtained with small $\alpha$ when $c_{sr}$ is relatively large. {We note that the PSNR performance without relay is $27.16$ dB which is much lower than that in Fig. \ref{fig:hd_alpha}.} 

Then, we compare the performance of the proposed DeepJSCC-PF with its AF counterpart as well as the digital baseline in Fig. \ref{fig:hd_pf_vs_bpg}. In this simulation, we set $P = 3$ dB for all the schemes. All the models with DeepJSCC-AF protocol adopt a fixed $\alpha = 1/2$ while the models with DeepJSCC-PF protocol select the optimal $\alpha^*$ given the channel conditions, i.e., $c_{sr}$ and $c_{rd}$. Various channel conditions with both $c_{sr}$ and $c_{rd}$ chosen from $\{0, 10/3, 20/3, 10\}$ dB are evaluated, and the PSNR performances are shown in Fig. \ref{fig:hd_pf_vs_bpg}. We also mark the optimal $\alpha^*$ on the plot for all the $16$ combinations of the DeepJSCC-PF scheme. It can be seen that both the proposed DeepJSCC-AF and DeepJSCC-PF protocols outperform the digital baseline by a large margin. Moreover, the DeepJSCC-PF outperforms the DeepJSCC-AF protocol in all the considered scenarios, which is intuitive as the DNNs at the relay should perform at least as well as linear scaling of the DeepJSCC-AF.  {We also notice that the DeepJSCC-PF outperforms its AF counterpart especially when the channel qualities $(c_{sr}, c_{rd})$ are poor. This might due to the fact that the neural network at the relay is not only capable to extract features from the received signal over the noisy $\mathrm{S}$-$\mathrm{R}$ link but also generates robust transmit signal $\bm{x}_{r}$ to combat the noise in the $\mathrm{R}$-$\mathrm{D}$ link.}  {Note that we provide additional 2d plots for the points which are invisible for all the 3d plots in this paper.}

\begin{table}[tbp]
\caption{{Evaluation of the important variables, $\alpha, \gamma$ and $\beta$ for the DeepJSCC-PF protocol with a half-duplex relay and $c_{rd} = 10/3$ dB, while $c_{sr} \in \{0, 10/3, 20/3, 10\}$ dB.}}
\begin{center}
\begin{tabular}{c|cccc}
\hline

\cline{1-4} 
\textbf{$c_{sr}$} (dB) & \textbf{0}& \textbf{10/3}& \textbf{20/3} & \textbf{10}\\
\hline

\textit{$\alpha$} &  4/6 & 4/6 & 3/6 & 2/6\\
\textit{$\gamma$} & 0.941 & 0.889 & 0.709 & 0.506\\
\textit{$\beta$} & 0.828 & 0.819 & 0.787 & 0.568\\

\hline
\end{tabular}
\label{tab:variables}
\end{center}

\end{table}

\subsubsection{Evaluations of important variables}\label{sec:variables}
We then evaluate the important variables, namely, the time division variable $\alpha$, power variable $\gamma$ and the correlation variable $\beta$, introduced in Section \ref{sec:half-duplex} to gain more insights regarding the proposed DeepJSCC-PF protocol. 

In this simulation, we adopt the same setting as in Fig. \ref{fig:hd_pf_vs_bpg}.
Note that the optimal $\alpha^*$ values (among the multiplies of $1/6$) for each link condition, $c_{sr}$ and $c_{rd}$ have already been marked in Fig. \ref{fig:hd_pf_vs_bpg}. We then explore the relationship of the variables, $\gamma$ and $\beta$ with respect to different link qualities. Due to the page limit, we provide the results for the following combinations, namely, $(c_{sr}, c_{rd}) = (0, 10/3)$, $(c_{sr}, c_{rd}) = (10/3, 10/3)$, $(c_{sr}, c_{rd}) = (20/3, 10/3)$ and $(c_{sr}, c_{rd}) = (10, 10/3)$ dB. As introduced in Section \ref{sec:half-duplex}, the calculation of $\gamma$ and $\beta$ follows $\gamma = \frac{\mathbb{E}{\|\bm{x}_s^{(1)}\|^2_2}}{\mathbb{E}{\|\bm{x}_s\|^2_2}}$ and $\beta = \frac{\mathbb{E}{(\bm{x}_r^\dagger \bm{x}_s^{(2)} })}{\sqrt{\mathbb{E}{\|\bm{x}_s^{(2)}\|^2_2} \mathbb{E}{\|\bm{x}_r\|^2_2}}}$, respectively.

As can be seen in Table \ref{tab:variables}, when $c_{sr}$ improves, the $\gamma$ of the learned models with the optimal $\alpha^*$ decreases. This confirms the intuition that, given a better $\mathrm{S}$-$\mathrm{R}$ link, the source can save its power in the \textit{relay-receive} period and consume more power in the \textit{relay-transmit} period for better performance. The correlation variable $\beta$ follows the same trend. 
We can argue that when the $\mathrm{S}$-$\mathrm{R}$ link is weak, the source mostly relies on direct transmission, allowing the relay to transmit only in the $1/3$ of the time and uses most of its power in the relay-receive period. On the other hand, what the source transmits in the relay-transmit period is highly correlated with relay's signal; that is, in this period, the main goal of the source is to collaborate with the relay to forward its received information. On the other hand, as the $\mathrm{S}$-$\mathrm{R}$ link quality improves, relay-receive period shortens, as the relay can quickly recovers the information from the source and uses the remaining time to forward it to the destination. We also see that the source uses increasing amount of its power in the relay-transmit period, and with a smaller $\beta$ value, which means that the source uses more of its power to transmit fresh information in this period, while also helping the relay forward its information to the destination.

\begin{table}[tbp]
\caption{{Comparison of the original DeepJSCC-PF protocol with the modified one under the same setting with Table \ref{tab:variables}. Note that the entries on the left correspond to the original protocol while those on the right to the modified one.}}
\begin{center}
\begin{tabular}{c|cccc}
\hline

\cline{1-4} 
\textbf{$c_{sr}$} (dB) & \textbf{0}& \textbf{10/3}& \textbf{20/3} & \textbf{10}\\
\hline

\textbf{PSNR} & \begin{tabular}{@{}c@{}}31.46$\pm 0.14$/ \\ 31.35$\pm 0.15$\end{tabular} & \begin{tabular}{@{}c@{}}32.18$\pm 0.14$/ \\ 32.10$\pm 0.15$\end{tabular} & \begin{tabular}{@{}c@{}}32.65$\pm 0.13$/ \\ 32.46$\pm 0.14$\end{tabular}  & \begin{tabular}{@{}c@{}}33.37$\pm 0.13$/ \\ 32.32$\pm 0.14$\end{tabular}  \\
\textit{$\gamma$} & 0.94/0.94 & 0.89/0.90 & 0.71/0.73 & 0.51/0.63\\
\textit{$\beta$} & 0.82/0.85 & 0.80/0.88 & 0.79/0.96 & 0.57/0.75\\

\hline
\end{tabular}
\label{tab:variables2}
\end{center}

\end{table}

\begin{figure}[t]
     \centering
     \begin{subfigure}{\columnwidth}
         \centering
         \includegraphics[width=0.7\columnwidth]{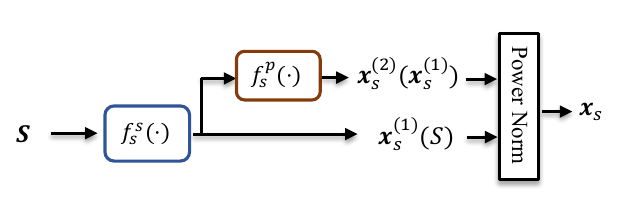}
         \caption{DeepJSCC-PF protocol.}
         \label{fig:protocol:a}
     \end{subfigure}     
     \begin{subfigure}{\columnwidth}
         \centering
         \includegraphics[width=0.7\columnwidth]{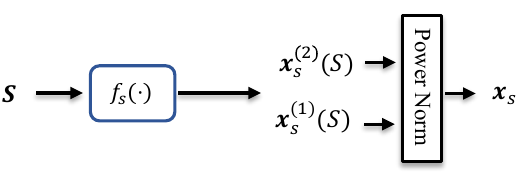}
         \caption{Modified protocol.}
         \label{fig:protocol:b}
     \end{subfigure}

  \caption{{Illustration of the source encoder for the DeepJSCC-PF protocol (a) and its modified version (b) where no new information is transmitted in the relay-transmit period.}}
  \label{fig:protocol}
\end{figure}

\subsubsection{{Transmission without new information}}
To better understand the source behaviour, we further consider an extreme case for the half-duplex relay mode, in which no new information is transmitted by the source node in the \textit{relay-transmit} period. As shown in Fig. \ref{fig:protocol:b}, we introduce the `systematic' ViT encoder at the source, denoted as $f_s^{s}(\cdot)$ to generate $\bm{x}_s^{(1)}$ with dimension $\alpha^*k$, then $\bm{x}_s^{(1)}$ is fed to the `parity' encoder, denoted as $f_s^{p}(\cdot)$, also a ViT encoder as in Fig. \ref{fig:vit_codec} but without the image-to-sequence transformation module  to obtain the `parity' signal $\bm{x}_s^{(2)}$ with dimension $(1-\alpha^*)k$. Note that $\bm{x}_s^{(2)}$ is generated using $\bm{x}_s^{(1)}$, so no new information can be transmitted by the source during the relay-transmit period. $\bm{x}_s^{(1)}$ and $\bm{x}_s^{(2)}$ are then power normalized to generate the final transmitted signal. 
To show the effect of transmitting the new information, in this experiment, we adopt the same setting with that in Table \ref{tab:variables} where the models are evaluated with varying $c_{sr} \in \{0, 10/3, 20/3, 10\}$ dB and a fixed $c_{rd} = 10/3$ dB.

We show the comparison between the original DeepJSCC-PF with the modified one in Table \ref{tab:variables2}. Note that we use $\alpha^*$ values reported in Table \ref{tab:variables}. As can be seen in the table, when the link qualities are bad, e.g., $c_{sr} = 0$ and $10/3$ dB, the original DeepJSCC-PF protocol outperforms the modified one by a small margin showing that in these cases, only a small amount of new information, is transmitted in the original DeepJSCC-PF protocol. 
When the $c_{sr}$ improves, it can be seen that transmitting new information is essential to achieve good performance where a 1 dB PSNR gain is observed when $c_{sr} = 10$ dB. It can also be seen from the table that the modified protocol always has a larger $\beta$ compared with the original one, 
as the source is not transmitting any new information it mainly tries to align its codeword with that of the relay for {beamforming} gains. The reason they cannot be aligned perfectly (unlike in the original DF scheme) is due to the noise in the $\mathrm{S}$-$\mathrm{R}$ link and the fact that the DeepJSCC-PF scheme cannot remove the noise in the received signal completely.

\subsection{Performance evaluation for full-duplex relaying} 
Next, we compare the relative performance of the DeepJSCC-AF, DeepJSCC-PF and the state-of-the-art BPG image compression algorithm delivered at a rate, denoted as $R^*_{fd}$, which is determined by the maximum of the achievable rates attained by the DF and CF protocols in the full-duplex case whose closed form expressions are given in the \cite{relay_capacity}. {Note that similar with the half-duplex scenario, the achievable rates of the DF and the CF protocols are also achieved via BMC and Wyner-Ziv coding, respectively.}

\begin{figure*}
\centering
\includegraphics[width=0.9\linewidth]{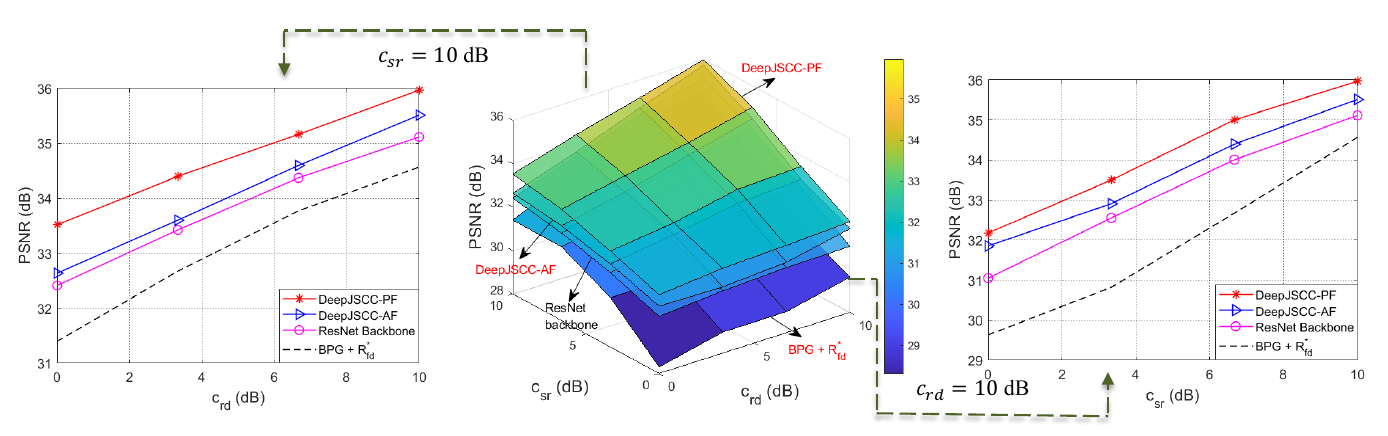}
\caption{{The PSNR performance of the proposed DeepJSCC-AF and PF protocols with ViT models and the digital baseline for the full-duplex relay channel with $c_{sr}, c_{rd} \in \{0, 10/3, 20/3 , 10\}$ dB. We also provide a benchmark of the DeepJSCC-PF protocol which uses ResNet as backbone.}}
\label{fig:fd_pf_vs_bpg}
\end{figure*}

\begin{table}[tbp]

    \caption{{The PSNR performance of the proposed DeepJSCC-PF protocol for the full-duplex relay channel with different number of blocks $B = \{2,3,4,6,8\}$ under $P = 3, c_{sr} = c_{rd} = 5$ dB.}}
\begin{center}
\begin{tabular}{c|ccccc}
\hline

\cline{1-4} 
\textbf{$B$} & \textbf{2}& \textbf{3}& \textbf{4} & \textbf{6} & \textbf{8}\\
\hline

\textit{PSNR} (dB)  &\begin{tabular}{@{}c@{}}32.06\\$\pm 0.12$ \end{tabular}  & \begin{tabular}{@{}c@{}}33.03\\$\pm 0.12$ \end{tabular} & \begin{tabular}{@{}c@{}}33.28\\$\pm 0.12$ \end{tabular} &  \begin{tabular}{@{}c@{}}33.53\\$\pm 0.12$ \end{tabular} &\begin{tabular}{@{}c@{}}33.62\\$\pm 0.11$ \end{tabular} \\

\hline
\end{tabular}
\label{tab:diff_B}
\end{center}

\end{table}

\subsubsection{DeepJSCC performance}
We consider the case with $B = 6$, $P = 3$ dB, while $c_{sr}$ and $c_{rd}$ are chosen from the set $\{0, 10/3, 20/3, 10\}$ dB. {The PSNR performance of the DeepJSCC-PF protocol with ResNet backbone is also provided where we set the number of hidden channels of the 2d-CNN layers to $C_{hid} = 256$ as in \cite{deepjsccq}. } The relative PSNR performance of the DeepJSCC schemes and the digital baseline are shown in Fig. \ref{fig:fd_pf_vs_bpg}.   As can be seen from the figure, all DeepJSCC schemes outperform the digital baseline. The full-duplex relay yields a better performance than its half-duplex counterpart, which is expected. Also aligned with our expectation, DeepJSCC-PF outperforms DeepJSCC-AF, but the gap between the two diminishes as $c_{sr}$ and $c_{rd}$ improve. This is due to the fact that less noise is forwarded by DeepJSCC-AF in this case. {Finally, it is observed that the ViT backbone outperforms its ResNet counterpart by $\sim 1$ dB.}

\subsubsection{Effects of different number of blocks}
Intuitively, there is a trade-off between the performance and the relay processing complexity (or, equivalently, the number of blocks). Note that when $B = 1$, the scheme degrades to point-to-point transmission while for $B = 2$, it degrades to a half-duplex relay since the relay keeps silent in the first block anyway. 

We set the number of blocks $B \in \{2, 3, 4, 6, 8\}$, $P = 3$ dB, $c_{sr} = c_{rd} = 5$ dB and present the achieved performances in Table \ref{tab:diff_B}. As can be seen, the reconstruction quality grows rapidly from $B = 2$ to $B = 6$, while it starts to saturate after $B = 8$. We set $B = 6$ for the simulations throughout this paper. It can also be verified in the figure that the half-duplex result matches the full-duplex case with $B = 2$.

\begin{table}[tbp]
    \caption{{Evaluation of the DeepJSCC-PF protocol for the full-duplex relay with different number of memory values $t \in \{1,2,5\}$ under $P = 3, c_{sr} = c_{rd} = 5$ dB and $B = 6$.}}
\begin{center}
\begin{tabular}{c|ccc|c}
\hline

\cline{1-4} 
\textbf{Memory $t$} & \textbf{1}& \textbf{2}& \textbf{5} & \textbf{DeepJSCC-AF}\\
\hline

\textit{PSNR (dB)} & \begin{tabular}{@{}c@{}}32.91\\$\pm 0.13$ \end{tabular} & \begin{tabular}{@{}c@{}}33.48\\$\pm 0.13$ \end{tabular} & \begin{tabular}{@{}c@{}}33.51\\$\pm  0.12$ \end{tabular} & \begin{tabular}{@{}c@{}}32.88\\$\pm  0.13$ \end{tabular}\\
\hline
\textit{SSIM} & \begin{tabular}{@{}c@{}}0.968\\$\pm  0.001$ \end{tabular} & \begin{tabular}{@{}c@{}}0.972\\$\pm  0.001$ \end{tabular} & \begin{tabular}{@{}c@{}}0.972\\$\pm  0.001$ \end{tabular} & \begin{tabular}{@{}c@{}}0.967\\$\pm  0.001$ \end{tabular}\\

\hline
\end{tabular}
\label{tab:memory}
\end{center}

\end{table}

\subsubsection{Relaying with memory}
As illustrated in Section \ref{sec:fd_relay}, in the conventional block Markov coding, the relay generates its transmitted signal for the $b$-th block, $\bm{x}_{r, b}$ based on $\bm{x}_{r,b-1}$ and $ \bm{y}_{r,b-1}$. In the DeepJSCC-PF, we generalize the operations at the relay in the BMC scheme by introducing the `knowledge matrix' $\bm{T}^{(b)}$, which includes all the information (not only the $(b-1)$-th block but also previous blocks) available so far to improve the reconstruction performance. 

To outline the effectiveness of the signal from previous blocks, $j<b$, we introduce an memory variable $t\in [1,B-1]$ which determines the maximum number of previous blocks used to generate the output $\bm{x}_{r,b}$. In particular, for the $b$-th block, the `knowledge matrix' $\bm{T}^{(b)}_t \in \mathbb{R}^{p^2\times \frac{2c^*(B-1)}{B}}$ with memory $t$ is given as
\begin{equation}
\bm{T}^{(b)}_t =
    \begin{cases}
    \begin{aligned}
        [\bm{0}, &\bm{Y}_{r, b-t}, \ldots, \bm{Y}_{r, b-1},\\ &  \bm{0}, \bm{X}_{r, b-t}, \ldots, \bm{X}_{r, b-1}],
    \end{aligned}
     & \text{if $b=B$.}\\
    \begin{aligned}
        [\bm{0}, &\bm{Y}_{r, b-t}, \ldots, \bm{Y}_{r, b-1}, \bm{0}, \\ &  \bm{0}, \bm{X}_{r, b-t}, \ldots, \bm{X}_{r, b-1}, \bm{0}],
    \end{aligned}
     & \text{if $t+1<b<B$.}\\
     \begin{aligned}
         [&\bm{Y}_{r, 1}, \ldots, \bm{Y}_{r, b-1}, \bm{0},\\ &\bm{X}_{r, 1}, \ldots, \bm{X}_{r, b-1}, \bm{0}],
     \end{aligned}
     & \text{otherwise.}
\end{cases}
\end{equation}

We present the performance of the DeepJSCC-PF with $B = 6$, $t \in \{1,2,5\}$ and $(c_{sr}, c_{rd}, P) = (5, 5, 3)$ dB in Table \ref{tab:memory}. Note that the $t = B-1 = 5$ case is the same with the DeepJSCC-PF curve in Fig. \ref{fig:fd_pf_vs_bpg}. It is shown in the table that having $t=1$ as in the conventional BMC scheme is not enough leading to a PSNR gap larger than $0.5$ dB compared with the $t=5$ case. {We also find that the DeepJSCC-AF protocol with a memory of one is only slightly outperformed by the DeepJSCC-PF with $t = 1$ indicating the gain of the DeepJSCC-PF protocol mainly comes from larger memory values.} Finally, it is observed that having $t=2$ is enough to achieve a similar reconstruction performance with that of $t=5$.

\begin{figure*}
\centering
\includegraphics[width=0.9\linewidth]{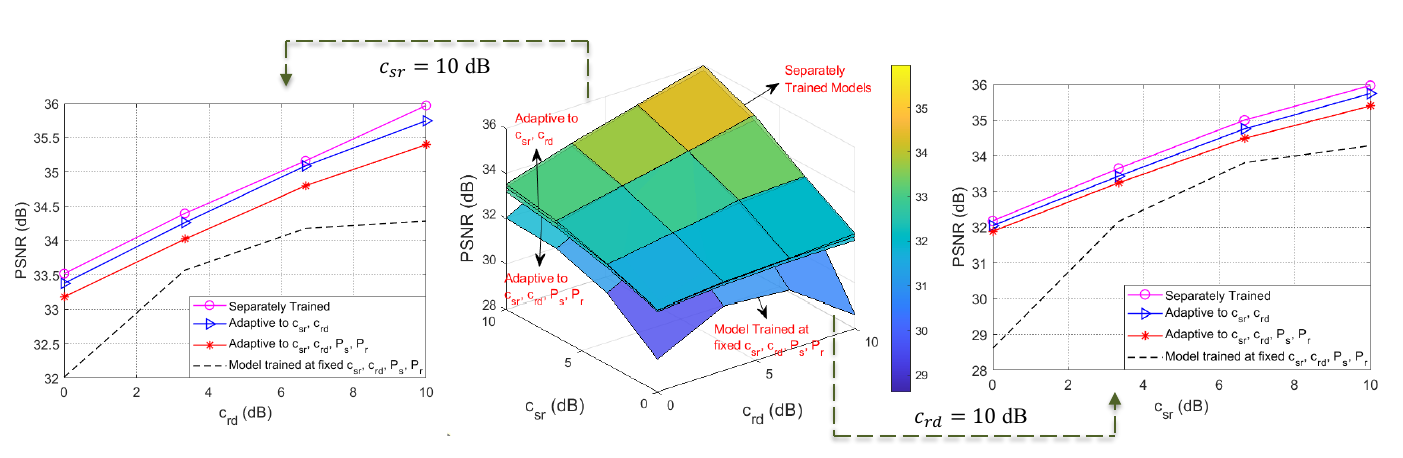}
\caption{{The PSNR comparison between the separately trained models (testing conditions are identical to the training conditions), the adaptive transmission models and the model trained at $c_{sr} = c_{rd} = 5$, $P_s = P_r = 3$ dB working in a full-duplex mode evaluated under $c_{sr}, c_{rd} \in \{0, 10/3, 20/3 , 10\}$ dB.}}
\label{fig:fd_adaptive}
\end{figure*}

  \begin{figure*}[t]
    \captionsetup[subfigure]{labelformat=empty}
    \captionsetup{font={normalsize}}
    \begin{subfigure}[b]{0.196\linewidth}
       \centering
       \caption{\textbf{Original}}
       \includegraphics[width=80pt]{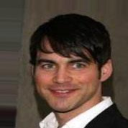}
       \vspace{-0.2cm}
       \caption{PSNR/SSIM}
    \end{subfigure}%
    \begin{subfigure}[b]{0.196\linewidth}
       \centering
       \caption{\textbf{BPG+$R_{hd}^{*}$}}
       \includegraphics[width=80pt]{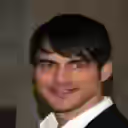}
       \vspace{-0.2cm}
       \caption{29.25/0.891}
    \end{subfigure}
    \begin{subfigure}[b]{0.196\linewidth}
       \centering
       \caption{\textbf{DeepJSCC-PF (HD)}}
       \includegraphics[width=80pt]{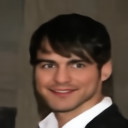}
       \vspace{-0.2cm}
       \caption{31.61/0.9370}
    \end{subfigure}
    \begin{subfigure}[b]{0.196\linewidth}
       \centering
       \caption{\textbf{BPG+$R_{fd}^{*}$}}
       \includegraphics[width=80pt]{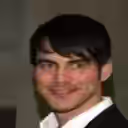}
       \vspace{-0.2cm}
       \caption{31.06/0.9177}
    \end{subfigure}
    \begin{subfigure}[b]{0.196\linewidth}
       \centering
       \caption{\textbf{DeepJSCC-PF (FD)}}
       \includegraphics[width=80pt]{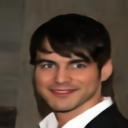}
       \vspace{-0.2cm}
       \caption{32.73/0.9501}
    \end{subfigure}

    \captionsetup[subfigure]{labelformat=empty}
    \captionsetup{font={normalsize}}
    \begin{subfigure}[b]{0.196\linewidth}
       \centering
       \includegraphics[width=80pt]{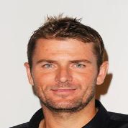}
       \vspace{-0.2cm}
       \caption{PSNR/SSIM}
    \end{subfigure}%
    \begin{subfigure}[b]{0.196\linewidth}
       \centering
       \includegraphics[width=80pt]{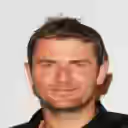}
       \vspace{-0.2cm}
       \caption{27.68/0.9236}
    \end{subfigure}
    \begin{subfigure}[b]{0.196\linewidth}
       \centering
       \includegraphics[width=80pt]{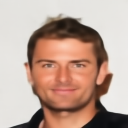}
       \vspace{-0.2cm}
       \caption{29.50/0.9579}
    \end{subfigure}
    \begin{subfigure}[b]{0.196\linewidth}
       \centering
       \includegraphics[width=80pt]{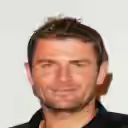}
       \vspace{-0.2cm}
       \caption{29.29/0.9470}
    \end{subfigure}
    \begin{subfigure}[b]{0.196\linewidth}
       \centering
       \includegraphics[width=80pt]{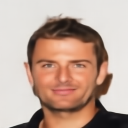}
       \vspace{-0.2cm}
       \caption{30.29/0.9640}
    \end{subfigure}
    
    \captionsetup[subfigure]{labelformat=empty}
    \captionsetup{font={normalsize}}
    \begin{subfigure}[b]{0.196\linewidth}
       \centering
       \includegraphics[width=80pt]{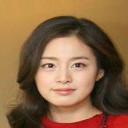}
       \vspace{-0.2cm}
       \caption{PSNR/SSIM}
    \end{subfigure}%
    \begin{subfigure}[b]{0.196\linewidth}
       \centering
       \includegraphics[width=80pt]{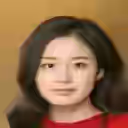}
       \vspace{-0.2cm}
       \caption{28.91/0.9374}
    \end{subfigure}
    \begin{subfigure}[b]{0.196\linewidth}
       \centering
       \includegraphics[width=80pt]{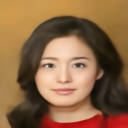}
       \vspace{-0.2cm}
       \caption{31.58/0.9644}
    \end{subfigure}
    \begin{subfigure}[b]{0.196\linewidth}
       \centering
       \includegraphics[width=80pt]{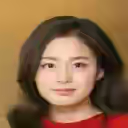}
       \vspace{-0.2cm}
       \caption{30.58/0.950}
    \end{subfigure}
    \begin{subfigure}[b]{0.196\linewidth}
       \centering
       \includegraphics[width=80pt]{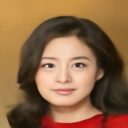}
       \vspace{-0.2cm}
       \caption{32.07/0.9692}
    \end{subfigure}

\captionsetup{font={small}}   
\caption{Visualizations of the reconstructed images from the CelebA dataset obtained by the proposed DeepJSCC-PF scheme as well as the BPG compression algorithm delivered at rates $R^*_{hd}$ and $R^*_{fd}$, respectively, with half-duplex and full-duplex relays, with $\text{CPP}=0.0234$. The PSNR and SSIM results are also provided for the reconstructed images.}
\label{fig:fig_result_celeba}
\end{figure*}

\subsubsection{Adaptive transmission model} Following the link quality adaptive model described in Section \ref{sec:adapt_trans}, we optimize two adaptive transmission models one with $(c_{min}, c_{max}) = (0, 10)$ dB {and $(P_{min}, P_{max}) = (0, 6)$ dB} while the other with $(c_{min}, c_{max}) = (0, 10)$ dB yet its transmit powers are fixed, i.e., $P_s = P_r = 3$ dB. In this simulation, we focus on the DeepJSCC-PF protocol\footnote{An adaptive transmission model can also be trained for the DeepJSCC-AF protocol, we will not show due to the page limit.} and adopt $B = 6$. The adaptively trained models are evaluated under {$c_{sr}, c_{rd} \in \{0, 10/3, 20/3, 10\}, P_s = P_r = 3$ dB}, whose overall reconstruction performance is compared with that of separately trained models in Fig. \ref{fig:fd_adaptive}. Since training an adaptive transmission model is generally harder, we adopt a slightly larger patience value, where the learning rate drops by $0.9$ if the validation loss does not improve in $25$ consecutive epochs ($20$ for the separately trained models). 
As can be seen from the figure, only a negligible PSNR gain is obtained by training distinct models for different network conditions, showing the effectiveness of the proposed `LA module' introduced in Section \ref{sec:fd_relay}. The model trained at a fixed network setting {$c_{sr} = c_{rd} = 5, P_s = P_r = 3$} dB achieves {slightly better reconstruction performance than the adaptive model when evaluated at the settings it is trained at}, but its performance degrades rapidly when the channel conditions change. Nevertheless, the curve obtained from the model trained at fixed {$c_{sr} = c_{rd} = 5, P_s = P_r = 3$} dB illustrates that the proposed DeepJSCC-PF protocol avoids the cliff and leveling effects as its PSNR performance gracefully degrades with lower $c_{sr}$ and $c_{rd}$ values while improving with better channel qualities.

\begin{figure}[t]
     \centering
     \begin{subfigure}{0.8\columnwidth}
         \centering
         \includegraphics[width=\columnwidth]{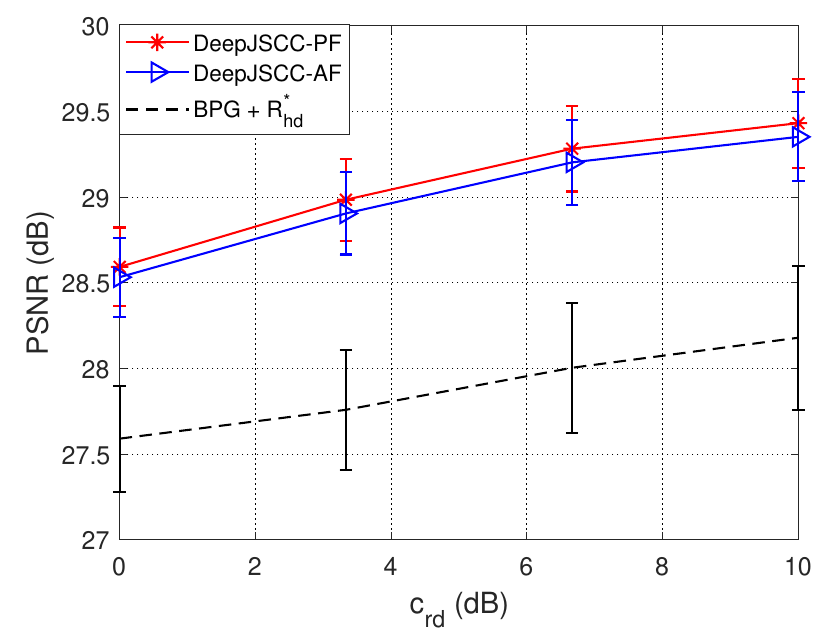}
         \caption{Half-duplex relaying with fixed $\alpha = 1/2$ for DeepJSCC-PF.}
     \end{subfigure}     
     \vspace{1cm}
     \begin{subfigure}{0.8\columnwidth}
         \centering
         \includegraphics[width=\columnwidth]{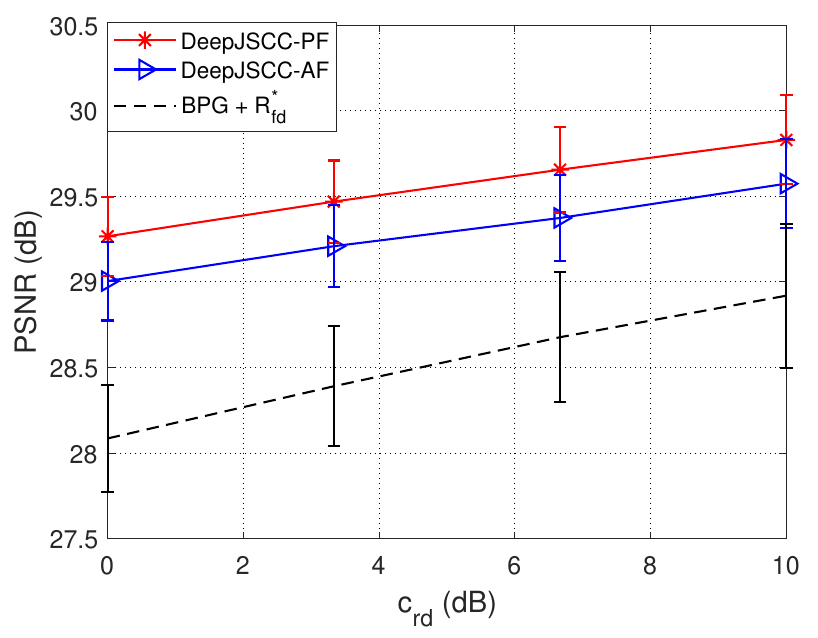}
         \caption{Full-duplex relaying.}
     \end{subfigure} 
  \caption{{PSNR performance obtained by the proposed DeepJSCC-PF, DeepJSC-AF and the separation baseline for both half and full-duplex relay over Rayleigh fading channel.}}
  \label{fig:fading_hdfd}
\end{figure}

{
\subsection{Fading channel}
We perform numerical experiments to compare the relative PSNR performance of the DeepJSCC-PF, DeepJSCC-AF and the BPG baseline under both half-duplex and full-duplex relaying modes over fading channel illustrated in Section \ref{sec:fading}. Note that the BPG baseline utilizes the same precoding algorithm as the proposed DeepJSCC schemes. In this simulation, we set $c_{sr} =  c_{sd} = 0$ dB, $c_{rd} = \{0, 10/3, 20/3, 10\}$ dB and $P_s = P_r = 3$ dB.  We first provide simulation results of the DeepJSCC-AF and DeepJSCC-PF protocols for the half-duplex relay. It is worth mentioning that, the optimal $\alpha$ for the DeepJSCC-PF protocol changes according to the random $h_{sr}, h_{rd}, h_{sd}$ realizations and the training algorithm illustrated in Algorithm \ref{algorithm:hd_pf} is no longer feasible. Thus, in this simulation, we simply fix the time-division variable, $\alpha = 1/2$ for the DeepJSCC-PF model and train it with different channel realizations. 
The relative performance of the proposed DeepJSCC schemes w.r.t the BPG baseline is shown in Fig. \ref{fig:fading_hdfd} (a). We can see the effectiveness of the proposed schemes as they outperform the BPG baseline by a large margin. It is also observed that the DeepJSCC-PF only outperforms the DeepJSCC-AF counterpart by a small margin due to a fixed $\alpha = 1/2$ value.
We then show the PSNR performance of the full-duplex relay  in Fig. \ref{fig:fading_hdfd} (b) where both DeepJSCC-AF and DeepJSCC-PF models outperform the BPG baseline.
}

{
\subsection{Complexity analysis}
In this subsection, we analyze the computational and time complexity of the proposed ViT models, and compare it with the BPG baseline. 

\textbf{Computational complexity}. We first analyze the computational complexity of the BPG algorithm which is implemented by first partitioning the image into blocks where each block has a fixed block size. With a slight abuse of notations,  we assume the block size equals to $B$. Then, discrete cosine transformation (DCT) is applied to each of the block followed by quantization and entropy coding to generate the bit sequence. The computational complexity of the block partitioning, quantization and entropy coding is neglectable and the total complexity is dominated by the DCT:
\begin{equation}
    O(\frac{M}{B^2}\times B^2\log B) = O(M \log B),
    \label{eq:complex_bpg}
\end{equation}
where $B^2\log B$ denotes the complexity of the DCT which is applied to each of the $\frac{M}{B^2}$ blocks with $M$ is the total number of pixels defined in \eqref{eq:psnr}.

The computational complexity of the ViT model is comprised of linear projections and the transformer layers shown in Fig. \ref{fig:vit_codec}. The number of tokens and the number of elements per token are denoted as $p^2$ and $N_t = \frac{M}{p^2}$, respectively. The first MLP layer requires $O(p^2 N_t c) = O(Mc)$ and the final MLP layer requires  $O({p^2} c c^*) = O(Mc\rho)$ where $c$ is the dimension of the hidden MLP layer, {$c^*= \frac{2\rho M}{p^2}$ and $\rho$ denotes the bandwidth ratio}. The complexity of each of the $N_e$ transformer layers is dominated by calculating the cross correlation between the tokens which can be expressed as $O(\frac{c^2M^2}{N_t^2})$. To sum up, the complexity of the ViT model is:
\begin{equation}
    O(Mc) + O(Mc\rho) + O\left(\frac{N_ec^2M^2}{N_t^2}\right) \approx O\left(\frac{N_e c^2M^2}{N_t^2}\right).
    \label{eq:complex_vit}
\end{equation}
By comparing \eqref{eq:complex_bpg} and \eqref{eq:complex_vit}, we can observe the complexity of the proposed ViT encoding is quadratic w.r.t the image size whereas the BPG scales linearly. This manifests that the proposed ViT encoding scheme achieves a superior performance at the cost of a higher computational complexity.

\textbf{Time complexity.}
We then show that the ViT model has lower time complexity. In particular, we run both the ViT and the BPG compression algorithm on the CIFAR10 test dataset which contains 5000 images. The encoding and decoding time of the two schemes are averaged over all the images which are shown in Table \ref{tab:complexity}. 

It can be seen that the proposed ViT scheme is much more efficient -- approximately ten times faster -- than the BPG baseline. This can be explained by the fact that the ViT encoding process does not need to compress the latent vectors into bits via entropy coding as in the BPG compression algorithm. Moreover, the ViT model can be accelerated using the modern GPU (we use RTX 4080 in this simulation) yet the BPG compression algorithm runs on the CPU.

\begin{table}[tbp]

    \caption{{Comparison of the computational and time complexity of the proposed ViT model and the BPG algorithm.}}
\begin{center}
\begin{tabular}{c|c|c}
\hline

 & \textbf{BPG}& \textbf{ViT} \\
\hline

\textit{Computational complexity} & $\mathcal{O}(M\log B)$  & $\mathcal{O}(\frac{N_e c^2 M^2}{N_t^2})$\\
\textit{Encoding/Decoding time (s)} & 0.05/0.13 & 0.003/0.005\\

\hline
\end{tabular}
\label{tab:complexity}
\end{center}

\end{table}
}

\subsection{Large datasets}
Finally, we evaluate our scheme on larger datasets and show that the learned DeepJSCC neural networks are capable to provide visually pleasing reconstructions in a large variety of network conditions. In this simulation, all the models are trained and tested with the CelebA \cite{celeba} dataset with a resolution of $128 \times 128$. To be precise, we consider the case where the link qualities are $(c_{sr}, c_{rd}, c_{sd}) = (10, 10, 0)$ dB while the transmission power at the source and the relay are set to $P = 3$ dB. The DeepJSCC-PF protocol for both the half-duplex and full-duplex scenarios is compared with the BPG image compression baseline. To be precise, the DeepJSCC-PF model for half-duplex relay is evaluated with $\alpha = 3/6$ (the best $\alpha$ value from the set $\{1/6, \cdots, 5/6\}$) while for the full-duplex relay, we set $B = 6$ and the memory size to $t = 5$. For the baseline scheme, we assume that the BPG compression output is delivered at the rates $R^*_{hd}$ and $R^*_{fd}$ for half-duplex and full-duplex cases, respectively.

It is worth mentioning that, for large datasets with high resolution, the ViT \cite{VIT} model used in the previous sections is no longer feasible. The reason is that, in the ViT model, the multi-head self attention is calculated between each token and all the remaining tokens, which leads to a quadratic complexity, $\mathcal{O}(M^2)$ with respect to the image size. When the image size is large, ViT models are less effective with high complexity. To solve this, researchers have proposed more advanced solutions, e.g., the Twins transformer \cite{twins} which is adopted as the backbone of the DeepJSCC models for the CelebA dataset and we refer interested readers to \cite{twins} for more details. For the simulation, the image-to-sequence transformation parameter is set to $p = 8$, which is identical to the configuration for the CIFAR-10 dataset and we have $c^* = 36$, resulting in a CPP value of $\rho = 0.0234$.

Visualizations of different CelebA images are provided in Fig. \ref{fig:fig_result_celeba}. As can be seen, for both half-duplex and full-duplex relays, the proposed DeepJSCC-PF outperforms the baseline separation-based scheme. In particular, DeepJSCC-PF not only yields superior PSNR and SSIM values but also produces more visually-pleasing reconstructions.

\section{Conclusion}
\label{sec:conclusion}
This paper presented a novel DeepJSCC scheme for image transmission over a cooperative relay channel, accommodating both the half-duplex and full-duplex relay scenarios {and is applicable to both AWGN and Rayleigh fading channels}. 
Our work presents two distinct DeepJSCC protocols, DeepJSCC-AF and DeepJSCC-PF. 
For enhanced adaptability in the context of full-duplex relay channels, we introduced the LA module, which allows a single DeepJSCC model to flexibly adjust its encoding to varying link qualities {and transmit powers}.
We demonstrated the efficacy of the proposed DeepJSCC-PF schemes through extensive numerical experiments using CIFAR-10 and CelebA datasets. Our results not only showcase impressive image reconstruction performance compared to the digital baseline, which employs the BPG compression algorithm and communicates at a rate achieved by the best of conventional DF and CF protocols, but also mitigate issues like the cliff and leveling effects in both half-duplex and full-duplex relay scenarios.

\bibliographystyle{IEEEbib}
\bibliography{refs}

\begin{IEEEbiography}[{\includegraphics[width=1.1in,height=1.3in,clip,keepaspectratio]{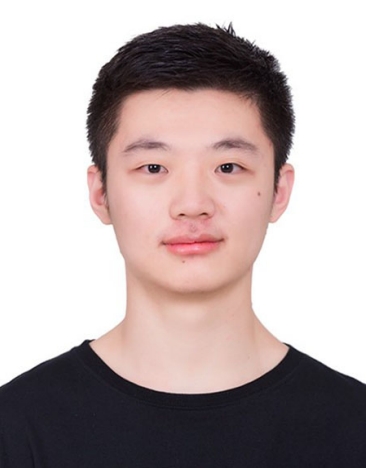}}]{Chenghong Bian} received the B.S. degree from the Physics Department, Tsinghua University, in 2020, and the M.S. degree from the EECS Department, University of Michigan, in 2022. He is currently pursuing the Ph.D. degree with the Department of Electrical and Electronic Engineering, Imperial College London. His research interests include wireless communications, machine learning and sensing. He received the Best Paper Award at IEEE International Conference on Communications 2023. 
\end{IEEEbiography}

\begin{IEEEbiography}[{\includegraphics[width=1.1in,height=1.3in,clip,keepaspectratio]{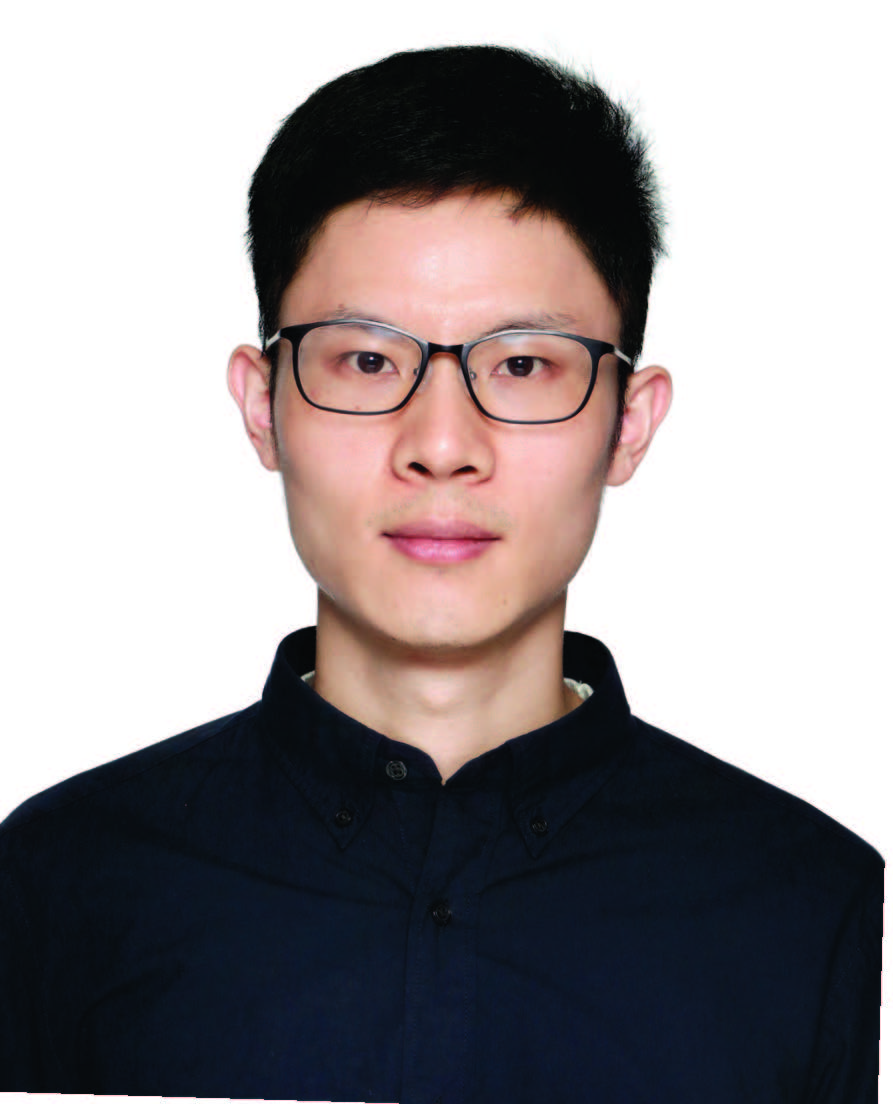}}]{Yulin Shao} (Member, IEEE) is an Assistant Professor with the State Key Laboratory of Internet of Things for Smart City, University of Macau, and a Visiting Researcher with the Department of Electrical and Electronic Engineering, Imperial College London. He received the B.S. and M.S. degrees in Communications and Information Engineering (Hons.) from Xidian University, China, in 2013 and 2016, and the Ph.D. degree in Information Engineering from the Chinese University of Hong Kong in 2020. He was a Research Assistant with the Institute of Network Coding, a Visiting Scholar with the Research Laboratory of Electronics at Massachusetts Institute of Technology, a Research Associate with the Department of Electrical and Electronic Engineering at Imperial College London, and a Lecturer in Information Processing with the University of Exeter. He was a Guest Lecturer at 5G Academy Italy and IEEE Information Theory Society Bangalore Chapter.

Dr. Shao's research interests include coding and modulation, machine learning, and stochastic control. He is a Series Editor of IEEE Communications Magazine in the area of Artificial Intelligence and Data Science for Communications, an Editor of IEEE Transactions on Communications in the area of Machine Learning and Communications, and an Editor of IEEE Communications Letters. He received the Best Paper Awards at IEEE International Conference on Communications (ICC) 2023, and IEEE Wireless Communications and Networking Conference (WCNC) 2024.
\end{IEEEbiography}

\begin{IEEEbiography}[{\includegraphics[width=1.1in,height=1.3in,clip,keepaspectratio]{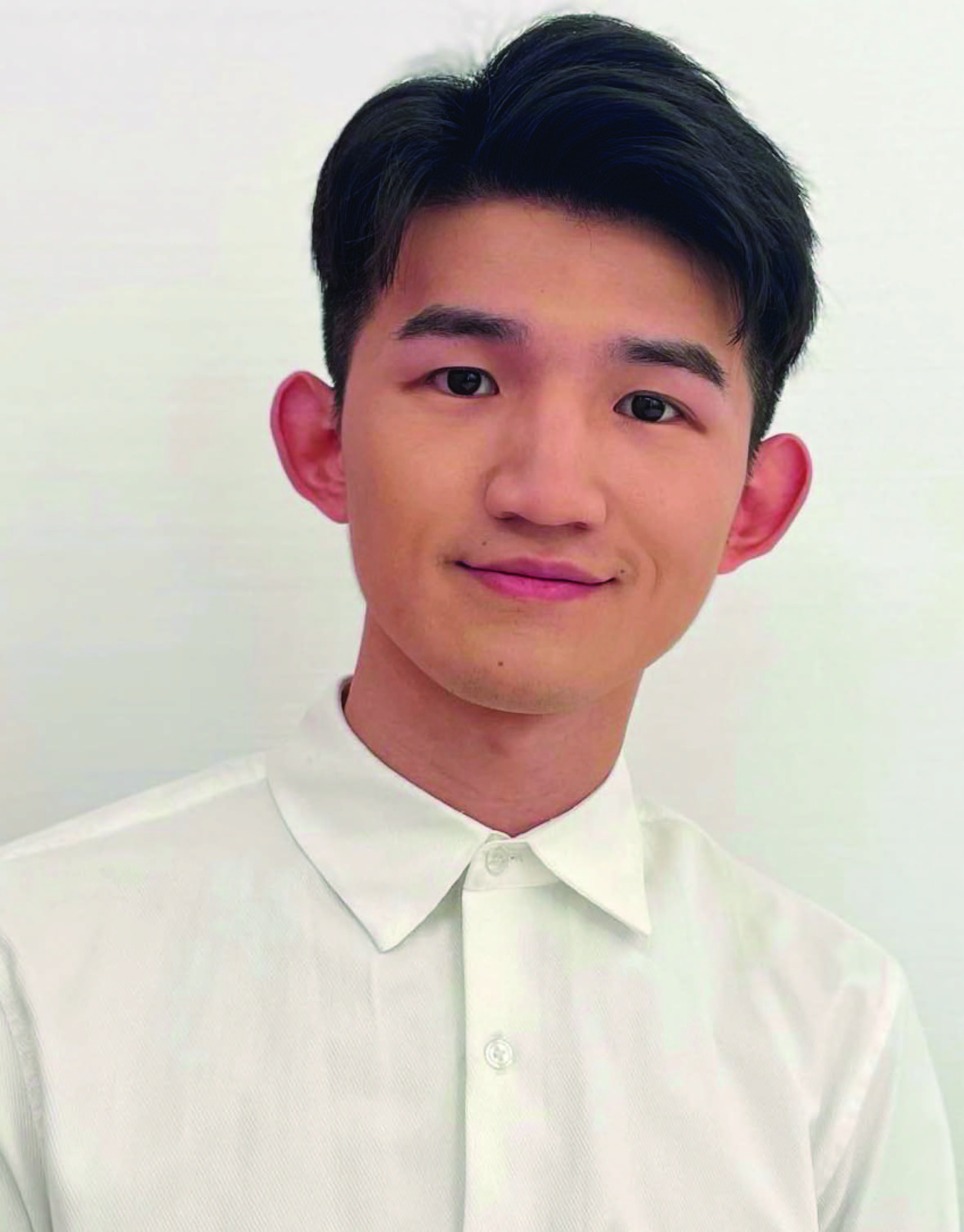}}]{Haotian Wu} (Graduate Student Member, IEEE) received the B.S. and M.S. degrees in Electrical Engineering from Zhejiang University, China, in 2017 and 2020, and a M.S. degree in Control Systems from Imperial College London in 2018. He is currently pursuing a Ph.D. degree at Imperial College London. His research interests include wireless communications and machine learning. He received the Best Paper Award at IEEE International Conference on Communications 2023. 
\end{IEEEbiography}

\begin{IEEEbiography}[{\includegraphics[width=1.1in,height=1.3in,clip,keepaspectratio]{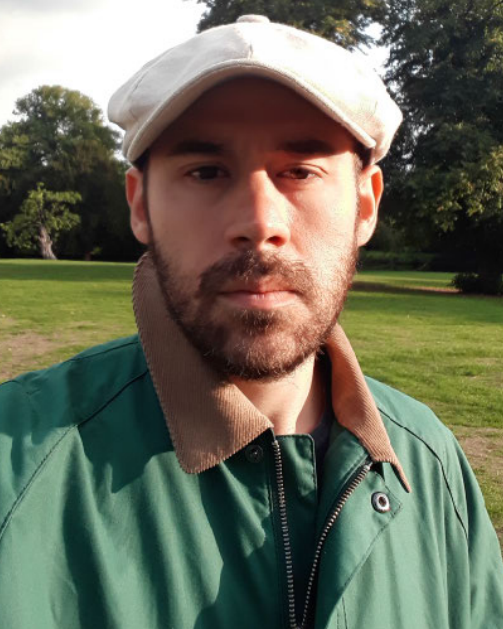}}]{Emre Ozfatura} received the B.Sc. degree in electronics engineering with a math minor and the M.Sc. degree in electronics engineering from Sabanci University, Turkey, in 2012 and 2015, respectively, and the Ph.D. degree from the Department of Electrical and Electronic Engineering, Imperial College London, U.K., in 2021. He is currently a Post-Doctoral Research Associate with the Information Processing and Communications (IPC) Laboratory, Imperial College London. His research interests include video streaming applications, distributed content storage, distributed computation, federated learning, and robust learning.
\end{IEEEbiography}

\begin{IEEEbiography}[{\includegraphics[width=1.1in,height=1.3in,clip,keepaspectratio]{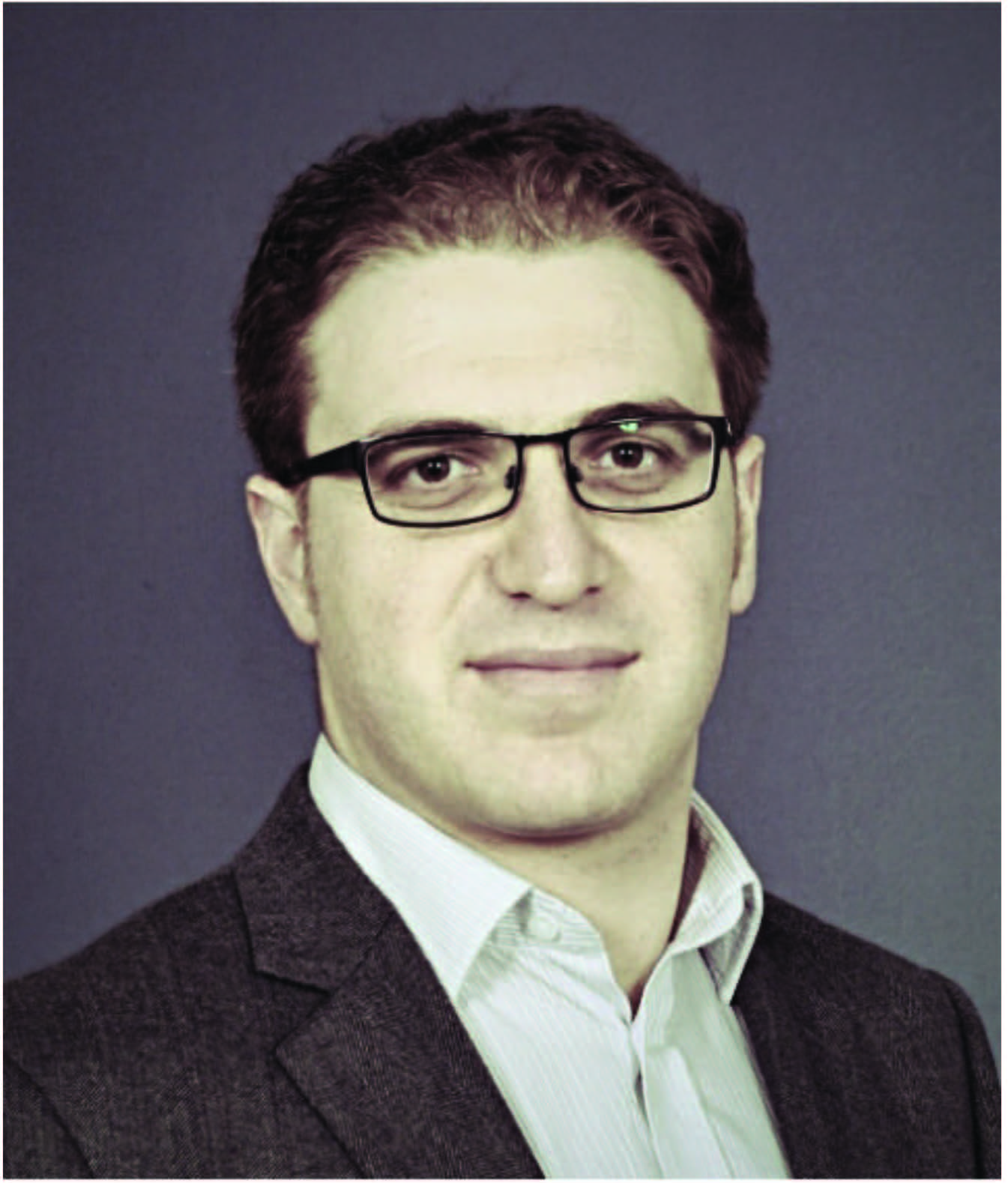}}]{Deniz~G\"und\"uz} (Fellow, IEEE) received the B.S. degree in electrical and electronics engineering from METU, Turkey in 2002, and the M.S. and Ph.D. degrees in electrical engineering from NYU Tandon School of Engineering (formerly Polytechnic University) in 2004 and 2007, respectively. Currently, he is a Professor of Information Processing in the Electrical and Electronic Engineering Department at Imperial College London, UK, where he also serves as the deputy head of the Intelligent Systems and Networks Group. In the past, he held various positions at the University of Modena and Reggio Emilia (part-time faculty member, 2019-22), University of Padova (visiting professor, 2018, 2020), Centre Tecnologic de Telecomunicacions de Catalunya (CTTC) (research associate, 2009-12), Princeton University (postdoctoral researcher, 2007-09, visiting researcher, 2009-11) and Stanford University (research assistant professor, 2007-09). His research interests lie in the areas of communications and information theory, machine learning, and privacy.

Deniz~G\"und\"uz is a Fellow of the IEEE. He is an elected member of the IEEE Signal Processing Society Signal Processing for Communications and Networking (SPCOM) and Machine Learning for Signal Processing (MLSP) Technical Committees. He serves as an Area Editor for the IEEE Transactions on Information Theory and IEEE Transactions on Communications. He is the recipient of the IEEE Communications Society Communication Theory Technical Committee (CTTC) Early Achievement Award in 2017, Starting (2016) and Consolidator (2022) and Proof-of-Concept (2023) Grants of the European Research Council (ERC), and has co-authored several award-winning papers, including the IEEE Communications Society - Young Author Best Paper Award (2022), and IEEE International Conference on Communications Best Paper Award (2023). He received the Imperial College London - President’s Award for Excellence in Research Supervision in 2023.
\end{IEEEbiography}

\end{document}